\documentclass[a4paper,11pt]{article}

\usepackage[margin=2.5cm]{geometry}

\usepackage[utf8]{inputenc}
\usepackage{microtype}
\usepackage{lmodern}
\usepackage{exscale}

\usepackage{amsmath}
\usepackage{amssymb}
\usepackage{mathtools}

\numberwithin{equation}{section}

\allowdisplaybreaks

\usepackage[linktocpage]{hyperref}
\usepackage[dvipsnames]{xcolor}
\hypersetup{
    colorlinks,
    linkcolor={purple},
    citecolor={purple},
    urlcolor={purple}
}

\definecolor{lightblue}{rgb}{0.737, 0.878, 0.918}
\definecolor{lightgreen}{rgb}{0.757, 0.984, 0.714}
\definecolor{lightred}{rgb}{1.000, 0.769, 0.804}
\definecolor{lightyellow}{rgb}{1.000, 0.914, 0.765}

\usepackage{cite}

\usepackage{mathrsfs}

\usepackage{relsize,exscale}
\usepackage{tensor}

\usepackage[low-sup]{subdepth}
\newcommand{\scr}{\scriptscriptstyle}

\usepackage[labelfont=bf]{caption}

\makeatletter
\newcommand{\dalembertian}{\mathop{\mathpalette\dalembertian@\relax}}
\newcommand{\dalembertian@}[2]{%
  \begingroup
  \sbox\z@{$\m@th#1\square$}%
  \dimen0=\fontdimen8
    \ifx#1\displaystyle\textfont\else
    \ifx#1\textstyle\textfont\else
    \ifx#1\scriptstyle\scriptfont\else
    \scriptscriptfont\fi\fi\fi3
  \makebox[\wd\z@]{%
    \hbox to \ht\z@{%
      \vrule width \dimen0
      \kern-\dimen0
      \vbox to \ht\z@{
        \hrule height \dimen0 width \ht\z@
        \vss
        \hrule height 2\dimen0
      }%
      \kern-2.5\dimen0
      \vrule width 2.5\dimen0
    }%
  }%
  \endgroup
}
\makeatother
\usepackage{orcidlink}

\begin{document}

\begin{center}

{\bf \Large \boldmath
Spectrum of pure~$R^2$ gravity: full Hamiltonian analysis
}

\

\renewcommand{\thefootnote}{\fnsymbol{footnote}}

{Will Barker}\,\orcidlink{0000-0002-1501-3221}\,\footnote{email: \href{mailto:barker@fzu.cz}{\tt barker@fzu.cz}}
{and Dra\v{z}en Glavan}\,\orcidlink{0000-0002-1983-0448}\,\footnote{email: \href{mailto:glavan@fzu.cz}{\tt glavan@fzu.cz}}

\setcounter{footnote}{0} 

\medskip

{\it CEICO, FZU --- Institute of Physics of the Czech Academy of Sciences,}
\\
{\it Na Slovance 1999/2, 182 21 Prague 8, Czech Republic}

\

\smallskip

\parbox{0.9\linewidth}{
We perform a full Hamiltonian constraint analysis of pure Ricci-scalar-squared ($R^2$) 
gravity to clarify recent controversies regarding its particle spectrum.
While it is well established  that the full theory consistently propagates three degrees of 
freedom, we confirm that its linearised spectrum 
around Minkowski spacetime is empty. Moreover, we show that this is not a feature unique 
to Minkowski spacetime, but a generic property of all 
traceless-Ricci spacetimes that have a vanishing Ricci
scalar, such as the Schwarzschild and Kerr black hole spacetimes. 
The mechanism for this phenomenon is a change in the nature of the constraints upon 
linearisation: 
ten second-class constraints of the full theory become first-class,
while the three momentum constraints degenerate into a single constraint. 
Furthermore, we show that higher order perturbation theory around these
singular backgrounds reveals no degrees of freedom at any order.
This is in conflict with the general analysis and points
to the fact that such backgrounds are surfaces of strong coupling in
field space, where the dynamics of perturbations becomes nonperturbative.
We further show via a cosmological phase-space analysis that the evolving universe is 
able to penetrate through the singular $R\!=\!0$ surface.
}

\end{center}

\

\hrule

\tableofcontents

%%%%%%%%%%%%%%%%%%%%%%%%%%%%%%%%%%%%%%%%%%%%%%%
%%%%%%%%%%%%%%%%%%%%%%%%%%%%%%%%%%%%%%%%%%%%%%%
%%%%%%%%%%%%%%%%%%%%%%%%%%%%%%%%%%%%%%%%%%%%%%%
%%%	INTRODUCTION
%%%%%%%%%%%%%%%%%%%%%%%%%%%%%%%%%%%%%%%%%%%%%%%
%%%%%%%%%%%%%%%%%%%%%%%%%%%%%%%%%%%%%%%%%%%%%%%
%%%%%%%%%%%%%%%%%%%%%%%%%%%%%%%%%%%%%%%%%%%%%%%
\section{Introduction}

%Recently some attention has been attracted by works reporting no propagating physical degrees of freedom in the linearised spectrum of perturbations in pure~$R^2$ theory around Minkowski space. This somewhat surprising, though correct result is seemingly at odds with the generally established result that~$f(R)$ theories --- that includes the pure~$R^2$ theory as a special case --- propagate three degrees of freedom: a graviton and a scalar.

Several recent works~\cite{Hell:2023mph,Golovnev:2023zen,Karananas:2024hoh} 
have pointed out that pure~$R^2$ gravity exhibits no physical propagating degrees of freedom 
in its linearised spectrum around Minkowski space. While this conclusion is 
ultimately correct, it is at first glance puzzling: the conventional wisdom is that generic~$f(R)$ theories --- 
of which pure~$R^2$ is a special case --- propagate three degrees of freedom: a graviton and a scalar.

%the discussion of this feature has unfolded principally among work by Hell, Lust and Zoupanos~\cite{Hell:2023mph}, Golovnev~\cite{Golovnev:2023zen} and Karananas~\cite{Karananas:2024hoh}. 

To our understanding, the strong coupling of the graviton around Minkowski space, and its absence from the
linearised spectrum was appreciated before~\cite{Alvarez-Gaume:2015rwa} (see also e.g.~\cite{DAmico:2020euu} 
and~\cite{Casado-Turrion:2023rni} that discuss the same issue). However, it was first pointed out in~\cite{Hell:2023mph}
that the scalar is also absent from the linearised spectrum, and that the spectrum of perturbative degrees of freedom 
around Minkowski space is empty. This work corrected the previous general analysis of quadratic 
gravity in~\cite{Alvarez-Gaume:2015rwa}, which reported a single scalar perturbative degree of freedom.
The feature of an empty spectrum was framed in terms of singular submanifolds in phase space and the subtleties of counting degrees 
of freedom in~\cite{Golovnev:2023zen}, which also attributed the origin of the discrepancy 
between~\cite{Hell:2023mph} and~\cite{Alvarez-Gaume:2015rwa} to an incorrect implementation of the Stueckelberg trick. 
The singular surface interpretation was confirmed in \cite{Karananas:2024hoh}, which identified an accidental gauge symmetry 
and associated strong coupling for linear perturbations around Minkowski space.\footnote{Since the exchange of articles 
in~\cite{Alvarez-Gaume:2015rwa,Hell:2023mph,Golovnev:2023zen,Karananas:2024hoh}, the question of what is propagating 
on Minkowski spacetime in any given tensorial field theory suddenly became very easy to answer, due to the availability of 
new software~\cite{Barker:2024juc}.}

While~\cite{Hell:2023mph,Golovnev:2023zen,Karananas:2024hoh} analysed
the linear spectrum of pure~$R^2$ theories around Minkowski space,
and explained the reasons for its emptiness, some aspects of this feature would still benefit 
from a better understanding. Firstly, the question of whether any other backgrounds might also exhibit 
an empty linearised spectrum was left open. 
Secondly, it remained unclear whether this feature survives at the nonlinear level;
it was suggested in~\cite{Hell:2023mph} that there might be no degrees of freedom around Minkowski space 
even at the nonlinear level, whereas~\cite{Karananas:2024hoh} pointed to the non-linear breakdown of the
accidental gauge symmetry behind emptying the linear spectrum. However, it should be noted that
both approaches utilized essentially perturbative methods, which might not be appropriate to describe
strongly coupled theories.
Ideally, the questions about  nonlinear aspects and background dependence
should be addressed by a full non-linear Hamiltonian analysis, which is what we set out to accomplish in this work.

An advantage of the Hamiltonian analysis is that it can be performed without any perturbative expansion whatever, 
and that in this non-perturbative setting any singular backgrounds are usually laid bare by some visible discontinuity 
in the constraint structure. Near such singular backgrounds, perturbation theory can still be \emph{attempted} in the 
Hamiltonian setting. There are, however, more and less consistent ways to do this. The approach in~\cite{Karananas:2024hoh} 
is to study the dynamics of truncated Taylor series, which seek to describe the $R^2$ action at varying degrees of fidelity. 
This procedure is successful in identifying the spectral discontinuity between quadratic and higher orders,
but it ignores the non-linear breakdown of diffeomorphism invariance induced by the truncation. A genuine 
perturbation theory, however, assumes the validity of the lowest-order approximation, to which higher-order corrections 
are added through an iterative procedure of substitutions, and this was not attempted in~\cite{Karananas:2024hoh}. Mechanistically, the iterative procedure can be implemented regardless of whether the lowest-order phenomena are correct or not. Catastrophic failure of the lowest-order approximation cannot be remedied by the addition of higher-order corrections, and this is a hallmark of \emph{non-perturbativity}~\cite{BeltranJimenez:2020lee}.

The Hamiltonian analysis of more general quadratic gravity theories has already been performed a long time ago in~\cite{Buchbinder:1987vp}, and our findings are completely in accord with theirs. However, we take a somewhat different route in performing the analysis, including a more convenient choice of canonical variables which allows for a transparent view of the subtleties arising in considering the theory perturbatively around particular backgrounds. It is especially the latter that allows us to better
understand the nature of the features reported in~\cite{Hell:2023mph,Golovnev:2023zen,Karananas:2024hoh}. 
We attribute the feature of disappearing degrees of freedom to the discontinuous change of the nature of constraints when considered perturbatively around specific backgrounds. Moreover, we establish this discontinuity not to be restricted to Minkowski spacetime, but a feature of {\it every} traceless-Ricci background,~$R\!=\!0$, such as Schwarzschild spacetime or Kerr spacetime, a possibility that was anticipated in~\cite{DAmico:2020euu,Casado-Turrion:2023rni,Golovnev:2023zen}.\footnote{Traceless Ricci spacetimes do not fall neatly into the Petrov classification~\cite{Petrov:2000bs}, which restricts only the Weyl tensor, but Pleba\'{n}ski provided the full four-dimensional classification~\cite{Plebanski:1964,McIntosh:1981}.} We also establish that considering non-linear perturbations around such backgrounds by perturbatively expanding the solutions will never recover any degrees of freedom. However, this is not a physical conclusion, but rather a limitation of the perturbative expansion that is unable to capture physical properties. The non-linear theory without this perturbative expansion reveals three propagating degrees of freedom in the vicinity of traceless-Ricci spacetimes.

Whilst the Hamiltonian analysis provides a necessary degree of closure, following the developments 
in~\cite{Hell:2023mph,Golovnev:2023zen,Karananas:2024hoh}, it also raises new questions. According to 
general lore, singular surfaces are separatrices: the phase space trajectories are not able to cross them or 
terminate on them (see e.g. the discussion in~\cite{BeltranJimenez:2020lee}). In equivalent terms, this lore 
states that strongly coupled backgrounds are dynamically unreachable by a propagation of the field equations 
from non-singular initial data. We investigate this principle for the cosmological sector of the pure $R^2$ theory, 
and are surprised to find that 
it does not seem to hold: a dynamical systems analysis allows a homogeneous universe to penetrate through 
the traceless-Ricci background at~$R\!=\!0$. This not only adds to the intrigue of the $R^2$ theory but, moreover, 
opens questions regarding the nature of strong coupling and singular surfaces in field space more generally.

We give the canonical formulation of the theory and its constraint analysis in Sec.~\ref{Hamiltonian analysis}. This is used in Sec.~\ref{sec: Perturbations around Minkowski space} to rederive the Minkowski space result for the linearised perturbations from the perspective of the full Hamiltonian analysis. We find that perturbing around the Minkowski background alters the nature of certain constraints: ten second-class constraints become first-class, while the momentum constraint becomes longitudinal, thereby removing two first-class constraints. These two features together result in the absence of propagating degrees of freedom in the linear spectrum. We also demonstrate how this feature is preserved in perturbation theory to higher orders, implying that the vicinity of Minkowski background should not be explored by perturbative methods. In Sec.~\ref{sec: Perturbations around other singular backgrounds}, we demonstrate that the strong coupling phenomenon identified in Minkowski space is, in fact, a general feature of all traceless-Ricci backgrounds within the pure~$R^2$ theory. The same change in the character of constraints observed for Minkowski space also appears in traceless-Ricci backgrounds, signaling the presence of accidental gauge symmetries, that we construct for Ricci-flat backgrounds in Appendix~\ref{sec: Perturbing Bach tensor}. We provide several explicit examples in Sec.~\ref{sec: Perturbations around other singular backgrounds} of other singular spacetimes, before proving the singularity of general traceless-Ricci backgrounds at linear and nonlinear levels, aided by technical details summarized in Appendix~\ref{sec: Higher order perturbations}. In Sec.~\ref{sec: Cosmological phase space} we address spacetimes that are not eternally traceless-Ricci, exploring whether their evolution can bring them through a traceless-Ricci phase. By analyzing the phase space of cosmological spacetimes within the pure~$R^2$ theory, we find that such a transition is indeed possible. This raises an intriguing question regarding the behavior of perturbations as the spacetime evolves through the singular surface. We further discuss this and related issues in Sec.~\ref{sec: Discussion}, including the potential for~$f(R)$ theories to exhibit analogous features.

%%%%%%%%%%%%%%%%%%%%%%%%%%%%%%%%%%%%%%%%%%%%%%%
%%%%%%%%%%%%%%%%%%%%%%%%%%%%%%%%%%%%%%%%%%%%%%%
%%%%%%%%%%%%%%%%%%%%%%%%%%%%%%%%%%%%%%%%%%%%%%%
%%%	HAMILTONIAN ANALYSIS
%%%%%%%%%%%%%%%%%%%%%%%%%%%%%%%%%%%%%%%%%%%%%%%
%%%%%%%%%%%%%%%%%%%%%%%%%%%%%%%%%%%%%%%%%%%%%%%
%%%%%%%%%%%%%%%%%%%%%%%%%%%%%%%%%%%%%%%%%%%%%%%
\section{Hamiltonian analysis}
\label{Hamiltonian analysis}

Determining the number of degrees of freedom in a given theory by examining the linear
spectrum of perturbations around a particular background is generally an unreliable method;
it only gives a lower bound on the number of degrees of freedom. The reliable answer is
provided by the full Hamiltonian constraint analysis~\cite{DiracBook}. In this section we 
perform a full Hamiltonian constraint analysis of the theory defined by the
action in~(\ref{R2action}) and obtain agreement with the analysis in~\cite{Buchbinder:1987vp}. 
This canonical approach allows for an unambiguous determination of the
physical degrees of freedom without relying on any perturbative expansion. The first step
is the Arnowitt-Deser-Misner (ADM) decomposition~\cite{Arnowitt:1962hi}, 
which foliates spacetime and isolates
the canonical phase space variables: the spatial metric, the extrinsic curvature, and their
conjugate momenta. Following this, the complete set of primary and secondary constraints
is systematically uncovered via the Dirac-Bergmann algorithm~\cite{Dirac:1950pj,Anderson:1951ta}, 
which requires that all
constraints be preserved under time evolution. The
resulting constraint algebra, formed by the Poisson brackets between all constraints, allows
for their definitive classification into first-class (possibly generating gauge symmetries) and
second-class (eliminating pairs of phase-space variables). This rigorous procedure not only
provides a definitive count of the propagating modes in the full theory but, as we will show
in Sec.~\ref{subsec: General traceless-Ricci spacetimes},
also precisely identifies the origin of the strong coupling pathology by revealing how the
constraint algebra itself degenerates on the singular~$R\!=\!0$ surfaces.

%%%%%%%%%%%%%%%%%%%%%%%%%%%%%%%%%%%%%%%%%%
%%%%%%%%%%%%%%%%%%%%%%%%%%%%%%%%%%%%%%%%%%
%%%	OVERVIEW OF THE THEORY
%%%%%%%%%%%%%%%%%%%%%%%%%%%%%%%%%%%%%%%%%%
%%%%%%%%%%%%%%%%%%%%%%%%%%%%%%%%%%%%%%%%%%
\subsection{Overview of the theory}
\label{subsec: Overview of the theory}

The pure~$R^2$ theory, a special case of a larger class 
of~$f(R)$ theories (see e.g.~\cite{DeFelice:2010aj,Nojiri:2010wj})
with~$f(R)\!=\!R^2$, is defined by its action,
\begin{equation}
S[g_{\mu\nu}] = \int\! d^4x \, \sqrt{-g} \, R^2 \, ,
\label{R2action}
\end{equation}
where the Ricci scalar,~$R=g^{\mu\nu}R_{\mu\nu}$, is the contraction of the Ricci
tensor,~$R_{\mu\nu} \!=\! \partial_\rho \Gamma^\rho_{\mu\nu} 
	\!-\! \partial_\nu \Gamma^\rho_{\rho\mu} 
	\!+\! \Gamma^\rho_{\mu\nu} \Gamma^\sigma_{\sigma\rho}
	\!-\! \Gamma^\rho_{\mu\sigma} \Gamma^\sigma_{\nu\rho}$, which in turn is defined
in terms of Christoffel symbols,~$\Gamma^\rho_{\mu\nu}
	\!=\! \frac{1}{2} g^{\rho\sigma} \bigl( \partial_\mu g_{\nu\sigma}
	\!+\! \partial_\nu g_{\mu\sigma}
	\!-\! \partial_\sigma g_{\mu\nu} \bigr)$.
Covariance is ensured by the presence of the metric determinant in the 
measure,~$g={\rm det}(g_{\mu\nu})$. 
Equations of motion for this theory are~\footnote{We should correct the remark
in~\cite{Hell:2023mph} about the pure~$R^2$ gravity possessing ``restricted Weyl symmetry'', i.e. 
being invariant under a local conformal 
rescaling of the metric, where the conformal factor is restricted to satisfy a covariant
Klein-Gordon equation. This theory, in fact, does not possess this property, 
since ``restricted Weyl symmetry'' is neither a local transformation, nor a symmetry,
as it does not leave equations of motion invariant~\cite{Glavan:2024svx}.} 
\begin{equation}
\Bigl( D_\mu D_\nu - g_{\mu\nu} D^\rho D_\rho
	- R_{\mu\nu} + \frac{1}{4} g_{\mu\nu} R \Bigr) R
	= 0 \, ,
\label{R2eom}
\end{equation}
where~$D_\mu$ denotes the covariant derivative compatible with the metric~$g_{\mu\nu}$. 
Note that spacetimes with a vanishing Ricci scalar,~$R=0$, automatically solve these 
equations. Such spacetimes are known as {\it traceless-Ricci} spacetimes, and will
play a central role in this work, following the Hamiltonian constraint analysis that this 
section is  devoted to. They include {\it Ricci-flat } spacetimes,~$R_{\mu\nu}=0$, as a 
special case that corresponds to vacuum solutions of Einstein's general relativity.

The theory in~(\ref{R2action}) is formulated in the Jordan frame, in which its
equations of motion~(\ref{R2eom}) contain higher derivatives of the metric. 
However,~$f(R)$ theories, including pure~$R^2$ as a special case, are more frequently 
considered in the Einstein frame, in which higher derivatives are traded for a scalar field, 
and where the metric dynamics is that of general relativity  coupled to the extra scalar. 
But the conformal transformation connecting these two frames is singular precisely 
for~$f'(R)=0$, which will turn out to be points of particular interest to us. Furthermore, 
Einstein frame~$f(R)$ theories, even though locally equivalent to their Jordan frame counterparts, 
are known not to be globally equivalent on account of this singularity in 
the transformation between them~\cite{Bahamonde:2016wmz,Alho:2016gzi,Rinaldi:2018qpu}.
For these reasons we refrain from considering the theory in the Einstein frame, and work 
directly in the Jordan frame throughout.

%%%%%%%%%%%%%%%%%%%%%%%%%%%%%%%%%%%%%%%%%%
%%%%%%%%%%%%%%%%%%%%%%%%%%%%%%%%%%%%%%%%%%
%%%	ADM DECOMPOSITION
%%%%%%%%%%%%%%%%%%%%%%%%%%%%%%%%%%%%%%%%%%
%%%%%%%%%%%%%%%%%%%%%%%%%%%%%%%%%%%%%%%%%%
\subsection{ADM decomposition}
\label{subsec: ADM decomposition}

The first step towards the canonical formulation of the theory in~(\ref{R2action}) is
the ADM decomposition~\cite{Arnowitt:1962hi} of the metric,
\begin{equation}
g^{00} = - \frac{1}{N^2} \, ,
\qquad \qquad
g_{0i} = N_i \, ,
\qquad \qquad
g_{ij} = h_{ij} \, ,
\label{ADMmetric1}
\end{equation}
where ADM variables are comprised of the lapse  scalar~$N$, the shift vector~$N_i$, 
and the spatial metric tensor~$h_{ij}$ induced on equal-time slices. The inverse 
components of the spacetime metric are then decomposed as,
\begin{equation}
g_{00} = - N^2 + N_i N^i \, ,
\qquad \qquad
g^{0i} = \frac{N^i}{N^2} \, ,
\qquad \qquad
g^{ij} = h^{ij} - \frac{N^i N^j}{N^2} \, .
\label{ADMmetric2}
\end{equation}
where~$h^{ij}$ is the inverse spatial metric,~$h_{ij} h^{jk} = \delta_i^k$,
that is henceforth used to raise indices on ADM variables, e.g.~$N^i = h^{ij} N_j$.
The metric determinant also has a simple ADM decomposition,~$\sqrt{-g} = N \sqrt{h}$.
Note that lapse is not allowed to vanish,~$N \!\neq\! 0$, in order to respect
invertibility of the metric~$g_{\mu\nu}$.

Apart from the metric we also need convenient variables for its time derivatives.
For the first time derivative this is provided by the extrinsic curvature tensor,
\begin{equation}
K_{ij} = - \frac{1}{2N} \Bigl( \dot{h}_{ij} - \nabla_i N_j - \nabla_j N_i \Bigr) \, .
\label{Kdef}
\end{equation}
where~$\nabla_i$ is the three-dimensional covariant derivative
with respect to the spatial metric~$h_{ij}$ and its corresponding Christoffel 
symbol,~$\gamma^{k}_{ij} = \frac{1}{2} h^{kl} \bigl( \partial_i h_{jl} + \partial_j h_{il} - \partial_l h_{ij} \bigr)$. 
Since pure~$R^2$ theory is a higher derivative theory, we also need a convenient variable for the 
second time derivative of the metric.
This is provided by a quantity introduced in~\cite{Buchbinder:1987vp},\footnote{We use a shifted
definition for~$F_{ij}$ compared to~\cite{Buchbinder:1987vp}, as we find it more convenient, 
but this is inessential.}
\begin{equation}
F_{i j} = - \frac{1}{N} \dot{K}_{ij}
	- K_{ik} {K^k}_j
	+ \frac{N^k}{N} \nabla_k K_{ij}
	+ \frac{2}{N} K_{k(i} \nabla_{j)} N^k
	- \frac{1}{N} \nabla_i \nabla_j N \, ,
\label{Fdef}
\end{equation}
with the shorthand notation for its contraction,~$F\!=\!h^{ij} F_{ij}$. 

After some work, the ADM decomposition of the Ricci scalar 
follows~\cite{Gourgoulhon:2007ue,Jha:2022svf},\footnote{Deriving ADM decomposition of various curvature 
tensors, and scalar invariants is greatly facilitated by the use of 
{\it Cadabra}~\cite{Peeters:2007wn,Peeters:2006kp,Peeters:2018dyg},
as was done in~\cite{Glavan:2024cfs}.}
\begin{equation}
R = 2F + K^2 - K^{ij} K_{ij} + \mathcal{R} \, ,
\label{Rdecomposition}
\end{equation}
where~$K \!=\! K^i{}_i$, and where~$\mathcal{R}\!=\! h^{ij} \mathcal{R}_{ij}$
is the Ricci scalar induced on spatial slices, that is formed as a contraction of the induced
Ricci tensor,
$\mathcal{R}_{ij} \!=\! \partial_k \gamma^k_{ij}
	\!-\! \partial_j \gamma^k_{k i} 
	\!+\! \gamma^k_{ij} \gamma^l_{l k}
	\!-\! \gamma^k_{i l} \gamma^l_{j k}$.
The
well-known identity in~(\ref{Rdecomposition}) finally allows us to write the 
action~(\ref{R2action}) in terms of ADM variables,
\begin{equation}
S \bigl[ N, N_i, h_{ij} \bigr]
	= 
	\int\! d^4 x \, N \sqrt{h} \, \Bigl( 2F + K^2 - K^{ij} K_{ij} + \mathcal{R} \Bigr)^{\!2}
	\, .
\label{ADMaction}
\end{equation}
%

%%%%%%%%%%%%%%%%%%%%%%%%%%%%%%%%%%%%%%%%%%
%%%%%%%%%%%%%%%%%%%%%%%%%%%%%%%%%%%%%%%%%%
%%%	CANONICAL ACTION
%%%%%%%%%%%%%%%%%%%%%%%%%%%%%%%%%%%%%%%%%%
%%%%%%%%%%%%%%%%%%%%%%%%%%%%%%%%%%%%%%%%%%
\subsection{Canonical action}
\label{subsec: Canonical action}

The action in~(\ref{ADMaction}) is still a higher derivative action, even though it is 
expressed in terms of ADM variables. In order to derive the first-order (i.e.~canonical)
formulation, we proceed by first constructing the extended action~\cite{Gitman}.
Here time derivatives are promoted to independent velocity fields,
\begin{equation}
K_{ij} \longrightarrow \mathcal{K}_{ij} \, ,
\qquad \qquad
F_{ij} \longrightarrow \mathcal{F}_{ij} \, ,
\end{equation}
and accompanying Lagrange multipliers~$\pi_{ij}$ and~$\rho_{ij}$ are
are introduced to ensure on-shell equivalence,
\begin{align}
\MoveEqLeft[2]
\mathcal{S} \bigl[ N, N_i, h_{ij}, \mathcal{K}_{ij}, \mathcal{F}_{ij}, \pi^{ij}, \rho^{ij} \bigr] 
	=
	\int\! d^4 x \, \biggl[
	N \sqrt{h} \, \Bigl( 2\mathcal{F} + \mathcal{K}^2 
		- \mathcal{K}^{ij} \mathcal{K}_{ij} + \mathcal{R} \Bigr)^{\!2}
	+ \pi^{ij} \Bigl( \dot{h}_{ij} - 2 \nabla_{(i} N_{j)} 
\nonumber \\
&
	+ 2 N \mathcal{K}_{ij} \Bigr)
	+ \rho^{ij} \Bigl( \dot{\mathcal{K}}_{ij}
		+ N \mathcal{K}_{ik} {\mathcal{K}^k}_j
		- N^k \nabla_k \mathcal{K}_{ij}
	- 2 \mathcal{K}_{k(i} \nabla_{j)} N^k
	+ \nabla_i \nabla_j N + N \mathcal{F}_{ij} \Bigr)
\biggr]
\, .
\label{ExtendedAction}
\end{align}
The canonical action is now constructed from the extended one above
by solving on-shell for as many components of~$\mathcal{F}_{ij}$ as possible, 
and plugging these back into the extended action~(\ref{ExtendedAction}) as off-shell equalities. 
Here it is possible to solve only for the trace,
\begin{align}
\MoveEqLeft[9]
\frac{ \delta \mathcal{S} }{ \delta \mathcal{F}_{ij} }
	=
	4 N \sqrt{h} \, \Bigl( 2\mathcal{F} + \mathcal{K}^2 
		- \mathcal{K}^{ij} \mathcal{K}_{ij} + \mathcal{R} \Bigr)
		h^{ij} + N \rho^{ij}
	\approx 0
\nonumber \\
&
\Longrightarrow \qquad 
\mathcal{F} \approx \overline{\mathcal{F}}
	=
	- \frac{1}{2} \Bigl( \mathcal{K}^2 
		- \mathcal{K}^{ij} \mathcal{K}_{ij} + \mathcal{R} \Bigr)
	- \frac{1}{24} \frac{\rho}{ \sqrt{h} }
	\, ,
\end{align}
while the transverse 
part~$\lambda_{ij} \!\equiv\! -\mathcal{F}_{ij} \!+\! \tfrac{1}{3} h_{ij} \mathcal{F}$
remains undetermined, and plays the role of the Lagrange multiplier. 
The canonical action is then written in the standard form,
\begin{align}
\MoveEqLeft[6]
\mathscr{S} \bigl[ N, N_i, \lambda_{ij}, h_{ij}, \pi^{ij}, \mathcal{K}_{ij}, \rho^{ij} \bigr] 
	\equiv
	\mathcal{S} \bigl[ N, N_i, h_{ij}, \mathcal{K}_{ij}, 
		\mathcal{F}_{ij}  \!\to\! \tfrac{1}{3} h_{ij} \overline{\mathcal{F}} \!-\! \lambda_{ij} , 
		\pi^{ij}, \rho^{ij} \bigr] 
\nonumber \\
={}&
	\int\! d^4x \, \Bigl[
		\pi^{ij} \dot{h}_{ij}
		+
		\rho^{ij} \dot{\mathcal{K}}_{ij}
		-
		N \bigl( \mathcal{H} + \lambda_{ij} \Phi^{ij} \bigr)
		-
		N_i \mathcal{H}^i
		\Bigr]
		\, ,
\label{CanonicalAction}
\end{align}
where the Hamiltonian and momentum constraints are, respectively,
\begin{align}
\mathcal{H}
	={}&
	\sqrt{h} \, \biggl[
	\frac{1}{144} \Bigl( \frac{\rho}{ \sqrt{h} } \Bigr)^{\!2}
	-
	2 \mathcal{K}_{ij} \frac{ \pi^{ij} }{ \sqrt{h} }
	+
	\frac{ 1 }{6} \Bigl( \mathcal{K}^2 - \mathcal{K}^{ij} \mathcal{K}_{ij} + \mathcal{R} \Bigr)
    		\frac{\mathcal{\rho} }{ \sqrt{h} }
	-
	\mathcal{K}_{ik} {\mathcal{K}^k}_j \frac{ \rho^{ij} }{ \sqrt{h} }
	-
	\nabla_i \nabla_j \Bigl( \frac{ \rho^{ij} }{ \sqrt{h} } \Bigr)
	\biggr]
	\, ,
\\
\mathcal{H}^i
	={}&
	\sqrt{h} \, \biggl[
	- 2 \nabla_j \Bigl( \frac{ \pi^{ij} }{ \sqrt{h} } \Bigr)
		+ \frac{\rho^{kl}}{ \sqrt{h} } \nabla^i \mathcal{K}_{kl}
		-
		2 \nabla^k \Bigl( \mathcal{K}^{il} \frac{ \rho_{kl} }{ \sqrt{h}} \Bigr)
	\biggr]
	\, ,
\end{align}
and where the primary traceless constraint,
\begin{equation}
\Phi^{ij} = \rho^{ij} - \frac{1}{3} h^{ij} \rho \, ,
\label{TracelessConstraint}
\end{equation}
appears multiplied by its traceless Lagrange multiplier~$\lambda_{ij}$.
Note that this multiplier can simplify the Hamiltonian and momentum constraints
by absorbing traceless parts of~$\rho^{ij}$,
\begin{align}
\mathcal{H} \longrightarrow{}&
	\sqrt{h} \biggl[
	\frac{1}{144} \Bigl( \frac{\rho}{ \sqrt{h} } \Bigr)^{\!2}
	-
	2 \mathcal{K}_{ij} \frac{ \pi^{ij} }{ \sqrt{h} }
	+
	\frac{1}{6} \Bigl( \mathcal{K}^2 - 3\mathcal{K}^{ij} \mathcal{K}_{ij} + \mathcal{R} \Bigr)  
		\frac{ \rho }{ \sqrt{h} }
	-
	\frac{1}{3}
	\nabla^i \nabla_i \Bigl( \frac{ \rho }{ \sqrt{h} } \Bigr)
	\biggr]
	\, ,
\label{HamiltonianConstraint}
\\
\mathcal{H}^i \longrightarrow{}&
	\sqrt{h} \biggl[
	- 2 \nabla_j \Bigl( \frac{ \pi^{ij} }{ \sqrt{h} } \Bigr)
	+
	\frac{1}{3} \frac{\rho}{ \sqrt{h} } \nabla^i \mathcal{K}
	-
	\frac{2}{3} \nabla_j \Bigl( \mathcal{K}^{ij} \frac{ \rho }{ \sqrt{h}} \Bigr)
	\biggr]
	\, .
\label{MomentumConstraint}
\end{align}
The action in~(\ref{CanonicalAction}) with the constraints in~(\ref{TracelessConstraint})--(\ref{MomentumConstraint})
is now the canonical formulation of the theory in~(\ref{R2action}). 

Varying the canonical action~(\ref{CanonicalAction}) with respect to variables~$h_{ij}$,
$\pi^{ij}$, $\mathcal{K}_{ij}$, and $\rho^{ij}$ generates their equations of 
motion,\footnote{Hamilton equations of motion have recently been derived for 
quadratic curvature theories in~\cite{Bellorin:2025kir}, 
but we cannot compare them to the ones derived here,
on account of the presence of the Ricci scalar-squared term, which precludes a smooth
limit in the canonical formulation.}
\begin{align}
\dot{h}_{ij} \approx{}&
	- 2 N \mathcal{K}_{ij}
	+ 2 \nabla_{(i} N_{j)}
	\, ,
\label{EOM1}
\\
\dot{\pi}^{ij}
	\approx{}&
	-
	\frac{N}{864} \frac{ \rho^2 }{ \sqrt{h} } h^{ij}
	-
	N \rho 
	\Bigl( h^{ik} h^{jl} - \frac{ 1 }{6} h^{ij} h^{kl} \Bigr)
	\Bigl( \mathcal{K}_{km} \mathcal{K}_{l}{}^m
		- \frac{1}{3} \mathcal{K}_{kl} \mathcal{K} \Bigr)
	+
	\frac{ N \mathcal{\rho} }{6} 
	\Bigl(
	\mathcal{R}^{ij}
	-
	\frac{ 1 }{3} \mathcal{R} h^{ij}
	\Bigr)
\nonumber \\
&
	-
	\frac{ 1 }{6} \sqrt{h} \, 
	\Bigl(
		N\nabla^i \nabla^j
		-
		Nh^{ij} \nabla^k \nabla_k
		-
		h^{ij} (\nabla^k N) \nabla_k \Bigr)
	\frac{\rho}{ \sqrt{h} }
	-
	\frac{ \rho }{6} \Bigl( \nabla^i \nabla^j - \frac{2}{3} h^{ij} \nabla^k \nabla_k \Bigr) N
\nonumber \\
&
	+
	\sqrt{h} \, \nabla_k
		\Bigl( N^k \frac{ \pi^{ij} }{ \sqrt{h} } - 2 N^{(i} \frac{ \pi^{j)k} }{ \sqrt{h}} \Bigr)
	+
	\frac{\rho}{3} \Bigl(
	N^k \nabla_k \mathcal{K}^{ij}
	+
	2\mathcal{K}^{k(i} \nabla^{j)} N_k
	-
	\frac{2}{3} h^{ij} \mathcal{K}^{kl} \nabla_k N_l
	\Bigr)
\nonumber \\
&
	-
	\frac{2}{3} \sqrt{h} \, N^{(i} \nabla_k
		\Bigl( \mathcal{K}^{j)k} \frac{\rho}{\sqrt{h}} \Bigr)
	+
	\frac{\rho}{3} 
		\Bigl( N^{(i} \nabla^{j)}
		-
		\frac{1}{3} h^{ij} N^k \nabla_k
		\Bigr)
		\mathcal{K}
	-
	\frac{N\rho}{3} \lambda^{ij}
	\, ,
\label{EOM2}
\\
\dot{\mathcal{K}}_{ij} \approx{}&
	\frac{h_{ij} }{3}
	\biggl[
	\frac{N}{2} \biggl(
	\frac{ 1 }{ 12 } \frac{\rho}{\sqrt{h}}
	+
	\mathcal{K}^2 
		\!-\! 3\mathcal{K}^{kl} \mathcal{K}_{kl} + \mathcal{R}
	\biggr)
	\!
	-
	\nabla_k \nabla^k N
	+
	N_k \nabla^k \mathcal{K}
	+
	2 \mathcal{K}_{kl} \nabla^k N^l
	\biggr]
	\!
	+
	N \lambda_{ij}
	\, ,
\label{EOM3}
\\
\dot{\rho}^{ij} \approx{}&
	2 N \pi^{ij}
	+
	N \rho\Bigl( \mathcal{K}^{ij} - \frac{ h^{ij}}{3} \mathcal{K} \Bigr)
	+
	\frac{h^{ij} }{3} \sqrt{h} \, \nabla_k \Bigl( N^k \frac{ \rho }{ \sqrt{h} } \Bigr)
	-
	\frac{2}{3} \rho \nabla_{(i} N_{j)}
	\, .
\label{EOM4}
\end{align}
These can be written in the form of Hamilton equations,
\begin{align}
\dot{h}_{ij} \approx \bigl\{ h_{ij} , H_{\rm tot} \bigr\} \, ,
\qquad
\dot{\pi}^{ij} \approx \bigl\{ \pi^{ij} , H_{\rm tot} \bigr\} \, ,
\qquad
\dot{\mathcal{K}}_{ij} \approx \bigl\{ \mathcal{K}_{ij} , H_{\rm tot} \bigr\} \, ,
\qquad
\dot{\rho}^{ij} \approx \bigl\{ \rho^{ij} , H_{\rm tot} \bigr\} \, ,
\end{align}
using the total Hamiltonian,
\begin{equation}
H_{\rm tot} = \int\! d^3x \, \Bigl[ N \bigl( \mathcal{H} + \lambda_{ij} \Phi^{ij} \bigr)
		+
		N_i \mathcal{H}^i \Bigr]
		\, ,
\end{equation}
and the canonical nonvanishing Poisson brackets,
\begin{equation}
\bigl\{ h_{ij}(t,\vec{x}) , \pi^{kl}(t,\vec{x}^{\,\prime}) \bigr\}
	=
	\delta_{(i}^k \delta^l_{j)} \delta^3(\vec{x} \!-\! \vec{x}^{\,\prime})
	\, ,
\qquad
\bigl\{ \mathcal{K}_{ij}(t,\vec{x}) , \mathcal{\rho}^{kl}(t,\vec{x}^{\,\prime}) \bigr\}
	=
	\delta_{(i}^k \delta^l_{j)} \delta^3(\vec{x} \!-\! \vec{x}^{\,\prime})
	\, ,
\end{equation}
encoded in the symplectic part of the action. Varying the action with respect to Lagrange 
multipliers~$N$,~$N_i$, and~$\lambda_{ij}$ generates ten primary constraints,
\begin{equation}
\mathcal{H} \approx 0 \, ,
\qquad \quad
\mathcal{H}^i \approx 0 \, ,
\qquad \quad
\Phi^{ij} \approx 0 \, .
\label{PrimaryConstraints}
\end{equation}
The conservation of these constraints as the system evolves in time is the focus of the following section.

%%%%%%%%%%%%%%%%%%%%%%%%%%%%%%%%%%%%%%%%%%%%%%%
%%%%%%%%%%%%%%%%%%%%%%%%%%%%%%%%%%%%%%%%%%%%%%%
%%%	CONSTRAINT ANALYSIS
%%%%%%%%%%%%%%%%%%%%%%%%%%%%%%%%%%%%%%%%%%%%%%%
%%%%%%%%%%%%%%%%%%%%%%%%%%%%%%%%%%%%%%%%%%%%%%%
\subsection{Constraint analysis}
\label{subsec: Constraint analysis}

Having derived the canonical formulation and identified all the primary 
constraints in~(\ref{PrimaryConstraints}), we proceed with the Dirac-Bergmann 
algorithm~\cite{Dirac:1950pj,Anderson:1951ta,DiracBook} for performing constraint 
analysis. This algorithm provides a reliable way of
counting the number~$N_{\rm phy}$ of propagating physical degrees of freedom,
corresponding to half the number of independent initial conditions needed to 
define the Cauchy problem. The algorithm requires the identification of all generations of 
constraints in order to identify the total number of first-class constraints~$N_{\rm 1st}$, and
second-class constraints~$N_{\rm 2nd}$. Then the number of physical degrees of freedom
is given by the formula
\begin{equation}
N_{\rm phy} = \frac{1}{2}
	\Bigl( N_{\rm can} - 2 N_{\rm 1st} - N_{\rm 2nd} \Bigr)
	\, ,
\label{Nphy}
\end{equation}
where~$N_{\rm can}$ is the number of canonical variables, not counting Lagrange 
multipliers.

Generations of constraints beyond the primary one are identified by requiring constraints
to be conserved. For systems without explicit time dependence
this is systematically inferred from the constraint algebra,
best presented in terms of smeared constraints,
\begin{equation}
\mathcal{H}[f] \equiv \int\! d^3x \, f(x) \mathcal{H}(x) \, ,
\qquad
\mathcal{H}^i[f_i] \equiv \int\! d^3x \, f_i(x) \mathcal{H}^i(x) \, ,
\qquad
\Phi^{ij}[f_{ij}] \equiv \int\! d^3x \, f_{ij}(x) \Phi^{ij}(x) \, .
\end{equation}
The smearing functions~$f$,~$f_i$, and~$f_{ij}$
in the definitions above  are strictly assumed to be independent of canonical variables, 
and consequently they have vanishing Poisson brackets with all quantities. 
%In the 
%remainder of this section we raise the indices of smearing functions with the  spatial
%metric,~e.g.~$f^i \!=\! h^{ij} f_j$, bearing in mind which index configuration depends
%<on the metric, and which one does not. 
We find the following on-shell algebra for primary 
constraints,
\begin{subequations}
\begin{align}
\bigl\{ \mathcal{H}[f] , \mathcal{H}[s] \bigr\}
	\approx{}&
	2 \Psi^{ij}\bigl[ s \nabla_i \nabla_j f \!-\! f \nabla_i \nabla_j s \bigr]
	\, ,
\label{bracket1}
\\
\bigl\{ \mathcal{H}[f] , \mathcal{H}^i[s_i] \bigr\}
	\approx{}&
	4 \Psi^{ij} \bigl[ f \mathcal{K}_{i}{}^k \nabla_j s_k \bigr]
	+
	2 \Psi^{ij} \bigl[ f s_k \nabla^k \mathcal{K}_{ij} \bigr]
	\, ,
\\
\bigl\{ \mathcal{H}^i[f_i] , \mathcal{H}^j[s_j] \bigr\}
	\approx{}&
	0
	\, ,
\\
\bigl\{ \Phi^{ij}[f_{ij}] , \mathcal{H}[s] \bigr\}
	\approx{}&
	2 \Psi^{ij}[s f_{ij}]
	\, ,
\\
\bigl\{ \Phi^{ij}[f_{ij}] , \mathcal{H}^k[s_k] \bigr\}
	\approx{}&
	0
	\, ,
\\
\bigl\{ \Phi^{ij}[f_{ij}] , \Phi^{kl}[s_{kl}] \bigr\}
	\approx{}&
	0
	\, ,
\label{bracket6}
\end{align}
\end{subequations}
where the nonvanishing quantity on the right-hand side is
\begin{equation}
\Psi^{ij}[f_{ij}] \equiv \int\! d^3x \, f_{ij}(x) \Psi^{ij}(x) \, ,
\qquad \text{where} \qquad
\Psi^{ij}(x) = \pi^{ij} - \frac{ 1 }{3} h^{ij} \pi
	+
	\Bigl(
	\mathcal{K}^{ij}
	-
	\frac{ 1 }{3} h^{ij} \mathcal{K}
	\Bigr)
	\frac{\rho}{6}
	\, .
\label{SecondaryConstraint}
\end{equation}
The conservation of primary constraints, according to the algebra above, necessitates
us to identify a secondary traceless constraint,
\begin{equation}
\Psi^{ij} \approx 0 \, ,
\label{SecondaryTracelessConstraint}
\end{equation}
which implies that all the brackets between primary constraints vanish.
Bracket of the secondary traceless constraint with itself vanishes, 
\begin{equation}
\bigl\{ \Psi^{ij}[f_{ij}] , \Psi^{kl}[s_{kl}] \bigr\}
	\approx
	0
	\, ,
\end{equation}
but one with the primary traceless constraint does not,
\begin{equation}
\bigl\{ \Phi^{ij}[f_{ij}]  , \Psi^{kl}[s_{kl}] \bigr\}
	\approx
	\int\! d^3x \,
	f_{ij} s_{kl}
	\Bigl(
	h^{i(k} h^{l)j}
	-
	\frac{1}{3} h^{ij} h^{kl}
	\Bigr)
	\frac{ \rho }{6}
	\, ,
\label{TracelessPoisson}
\end{equation}
except at singular points where~$\rho\approx 0$, that will be discussed in
the remainder of the paper.
Rather than generating further constraints, the conservation of the secondary traceless 
constraint determines the Lagrange multiplier on-shell,
\begin{align}
\lambda_{ij} \approx \overline{\lambda}_{ij}
	\equiv{}&
	\Bigl( \delta_{(i}^k \delta_{j)}^l - \frac{1}{3} h_{ij} h^{kl} \Bigr)
	\biggl[
	\mathcal{R}_{kl} 
	- 2 \mathcal{K}_k{}^m \mathcal{K}_{ml}
	+ \frac{2}{3} \mathcal{K} \mathcal{K}_{kl}
	- \frac{1}{\rho} \Bigl( 2 \pi \mathcal{K}_{kl} + \nabla_k \nabla_l \rho \Bigr)
\nonumber \\
&
	+ \frac{1}{N} \Bigl( 2 \mathcal{K}^{m(k} \nabla^{l)} N_m
		+ N^m \nabla_m \mathcal{K}_{kl}
		- \nabla_k \nabla_l N \Bigr)
	\biggr]
	\, .
\label{LambdaOnShell}
\end{align}
The fact that this quantity does not vanish means that the Poisson brackets between the 
primary Hamiltonian and momentum constraints in~(\ref{HamiltonianConstraint})
and~(\ref{MomentumConstraint})
with the secondary traceless constraint 
do not vanish. Superficially this might seem as though all constraints are second-class. 
However, that this is not the case is revealed by 
shifting the off-shell Lagrange multiplier by the value~(\ref{LambdaOnShell})
it takes on-shell,
\begin{equation}
\lambda_{ij} \longrightarrow \lambda_{ij} + \overline{\lambda}_{ij} \, .
\end{equation}
This changes the on-shell value of the multiplier to zero,~$\lambda_{ij}\!\approx\!0$,
and modifies the Hamiltonian and momentum constraints.
\begin{align}
&
\mathcal{H} +
	\Bigl[ 
	\mathcal{R}_{ij}
	- 2 \mathcal{K}_i{}^k \mathcal{K}_{kj}
	+ \frac{2}{3} \mathcal{K} \mathcal{K}_{ij}
	- \frac{1}{\rho} \Bigl( 2 \pi \mathcal{K}_{ij} + \nabla_i \nabla_j \rho \Bigr)
	- \nabla_i \nabla_j \Bigr]\Phi^{ij}  
	\longrightarrow
	\mathcal{H}
	\, ,
\\
&
\mathcal{H}^i
	- 2\nabla^k \bigl( \mathcal{K}^{ij} \Phi_{jk} \bigr)
	+ \Phi^{jk} \nabla^i \mathcal{K}_{jk}
	\longrightarrow
	\mathcal{H}^i
	\, .
\end{align}
In general, shifts of Lagrange multipliers effectively define different linear combinations
of primary constraints.\footnote{The change in the Hamiltonian and momentum 
constraints induced by shifting the Lagrange multiplier associated to another constraint 
is sometimes called {\it dressing} the Hamiltonian and momentum constraint, and
is necessary to correctly identify the first-class constraints, see e.g.~\cite{Alexandrov:2021qry}.}
This shift in the Hamiltonian and momentum constraint does not change anything
about the vanishing brackets in~(\ref{bracket1})--(\ref{bracket6}), but it makes the
brackets with the secondary traceless constraint vanish,
\begin{equation}
\bigl\{ \Psi^{ij}[f_{ij}] , \mathcal{H}[s] \bigr\} \approx 0 \, ,
\qquad \quad
\bigl\{ \Psi^{ij}[f_{ij}] , \mathcal{H}^k[s_k] \bigr\} \approx 0 \, .
\end{equation}
thereby identifying linear combinations of primary constraints that are
first-class.
Thus, the total number of first-class constraints 
is~$N_{\rm 1st}\!=\!4$, and the total number of second class constraints 
is~$N_{\rm 2nd} \!=\! 10$. Given that the number of canonical variables 
is~$N_{\rm can} \!=\! 24$, the number of physical degrees of freedom is
\begin{equation}
N_{\rm phy} = \frac{1}{2}
	\Bigl( 24 - 2 \!\times\! 4 - 10 \Bigr)
	=
	3
	\, ,
\end{equation}
which is the result obtained in~\cite{Buchbinder:1987vp}.

It should be noted that the number of physical propagating degrees of freedom does 
not depend on the choice of field variables, as long as field redefinitions are 
invertible.\footnote{Strictly speaking transformations should be invertible and non-singular,
as invertible singular transformations can introduce new degrees of 
freedom~\cite{Jirousek:2022jhh}.}
This includes shifting the field variables,
\begin{align}
&
h_{ij} \longrightarrow \overline{h}_{ij} + \delta h_{ij} \, ,
&&
\pi^{ij}\longrightarrow \overline{\pi}{}^{ij} + \delta \pi^{ij} \, ,
&&
\mathcal{K}_{ij} \longrightarrow \overline{\mathcal{K}}_{ij} + \delta \mathcal{K}_{ij} \, ,
&&
\rho^{ij} \longrightarrow \overline{\rho}{}^{ij} + \delta \rho^{ij} \, ,
\nonumber\\
&
N \longrightarrow \overline{N} + \delta N \, ,
&&
N_i \longrightarrow \overline{N}_i + \delta N_i \, ,
&&
\lambda_{ij} \longrightarrow \overline{\lambda}_{ij} + \delta \lambda_{ij} \, .
\label{BackgroundSplit}
\end{align}
In particular this is true for choosing the barred quantities to be solutions of the equations
of motion~(\ref{EOM1})--(\ref{EOM4}) and constraints~(\ref{PrimaryConstraints}).
In that case the new dynamical fields are understood to be perturbations around 
the background given by barred quantities. As long as we do not truncate the 
action for these perturbations and keep all of the terms, this shift
does not change the number of degrees of freedom.
However, truncating the action might easily do just that. 
This is in fact what happens in Minkowski space, that we examine in
Sec.~\ref{sec: Perturbations around Minkowski space}, and other critical spacetimes
we discuss in Sec.~\ref{sec: Perturbations around other singular backgrounds}.

Another aspect we should comment on is the step of reducing the phase space, that is in 
principle available after identifying and classifying all the constraints. This is accomplished
by using the second-class constraints to eliminate some canonical variables off-shell,
which is generally a legitimate step that reduces the dimensionality of phase space without 
changing physics. However, this step is not strictly necessary, and we refrain from taking it 
because of the subtlety with singular points in the bracket~(\ref{TracelessPoisson}).
We anticipate that this explicit reduction of phase space would introduce similar issues that 
the transformation between Jordan and Einstein frame does, that would interfere with 
our analysis of the singular points in the remainder of the paper.

%%%%%%%%%%%%%%%%%%%%%%%%%%%%%%%%%%%%%%%%%%%%%%%
%%%%%%%%%%%%%%%%%%%%%%%%%%%%%%%%%%%%%%%%%%%%%%%
%%%%%%%%%%%%%%%%%%%%%%%%%%%%%%%%%%%%%%%%%%%%%%%
%%%	PERTURBATIONS AROUND MINKOWSKI SPACE
%%%%%%%%%%%%%%%%%%%%%%%%%%%%%%%%%%%%%%%%%%%%%%%
%%%%%%%%%%%%%%%%%%%%%%%%%%%%%%%%%%%%%%%%%%%%%%%
%%%%%%%%%%%%%%%%%%%%%%%%%%%%%%%%%%%%%%%%%%%%%%%
\section{Perturbations around Minkowski space}
\label{sec: Perturbations around Minkowski space}

In this section we examine the degrees of freedom for perturbations
around Minkowski space, and how they are embedded into
the general picture of the Hamiltonian analysis of the preceding section. These 
perturbations are defined by shifts~(\ref{BackgroundSplit}) such that the background
field values are those of Minkowski space, for which the only nonvanishing ones are
\begin{equation}
\overline{h}_{ij} = \delta_{ij} \, ,
\qquad \quad
\overline{N} = 1 \, .
\end{equation}
We first establish that {\it linearised} perturbations, i.e.~perturbations defined
by truncating the action at quadratic order after the shift in~(\ref{BackgroundSplit}),
exhibit an empty spectrum of degrees of freedom. This is because linear truncation
changes the character of second-class constraints in the full theory to first-class, 
and furthermore removes the transverse part of the momentum constraint. 

We then proceed to examine higher order perturbation theory, finding consistently 
an empty spectrum to an arbitrary order. Nonetheless,
the correct interpretation of this observation is that perturbative expansion is
not an appropriate method for probing the vicinity of Minkowski space, where three 
degrees of freedom propagate in a full theory.

%%%%%%%%%%%%%%%%%%%%%%%%%%%%%%%%%%%%%%%%%%
%%%%%%%%%%%%%%%%%%%%%%%%%%%%%%%%%%%%%%%%%%
%%%	LINEAR PERTURBATIONS
%%%%%%%%%%%%%%%%%%%%%%%%%%%%%%%%%%%%%%%%%%
%%%%%%%%%%%%%%%%%%%%%%%%%%%%%%%%%%%%%%%%%%
\subsection{Linear perturbations}
\label{subsec: Linear perturbations}

Linear perturbations around Minkowski space are given by the quadratic canonical action,
\begin{align}
\MoveEqLeft[5]
\mathscr{S}_{\scr (2)}
	\bigl[ \delta N, \delta N_i, \delta \lambda_{ij}, \delta h_{ij}, \delta \pi^{ij}, 
		\delta \mathcal{K}_{ij}, \delta \rho^{ij} \bigr] 
\nonumber \\
={}&
	\int\! d^4x \, \Bigl[
		\delta \pi^{ij} \delta \dot{h}_{ij}
		+
		\delta \rho^{ij} \delta \dot{\mathcal{K}}_{ij}
		-
		%\mathscr{H}_{\scr (2)}
		\mathcal{H}_{\scr (2)}
		-
		\delta N \mathcal{H}_{\scr (1)}
		-
		\delta N_i \mathcal{H}^i_{\scr (1)}
		-
		\delta \lambda_{ij} \Phi^{ij}_{\scr (1)}
		\Bigr]
		\, .
\label{FirstOrderAction}
\end{align}
with linearised primary constraints given by
\begin{equation}
\mathcal{H}_{\scr (1)} =
	-
	\frac{1}{3}
	\partial_i \partial^i \delta \rho
	\, ,
\qquad \quad
\mathcal{H}^i_{\scr (1)} =
	- 2 \partial_j \delta \pi^{ij}
	\, ,
\qquad \quad
\Phi^{ij}_{\scr (1)}
	=
	\delta\rho^{ij} 
	-
	\frac{1}{3} \delta^{ij} \delta \rho \, ,
\label{LinearMinkConstraints}
\end{equation}
where now indices are raised and lowered by the Kronecker delta symbol.
The canonical Hamiltonian density for linearised perturbations
receives contributions only from the second perturbation of the Hamiltonian constraint,
\begin{equation}
\mathcal{H}_{\scr (2)}
	=
	\frac{ \delta\rho^2 }{144}
	-
	2 \delta \mathcal{K}_{ij} \delta \pi^{ij}
	+
	\frac{ \delta\rho }{6}
		\bigl( \partial^i \partial^j - \delta^{ij} \partial^k \partial_k  \bigr) \delta h_{ij}
	\, .
\label{Mink1stH}
\end{equation}

Requiring the conservation of primary constraints,
\begin{equation}
\dot{\mathcal{H}}_{\scr (1)}
	\approx
	- \frac{2}{3} \partial^i \partial_i \delta\pi
	\approx
	2 \partial_i \partial_j \Psi_{\scr (1)}^{ij} \, ,
\qquad\quad
\dot{\mathcal{H}}^i_{\scr (1)}
	\approx
	0 \, ,
\qquad\quad
\dot{\Phi}_{\scr (1)}^{ij}
	\approx
	2 \Psi_{\scr (1)}^{ij} \, ,
\end{equation}
now generates the linearised secondary traceless constraint,
\begin{equation}
\Psi^{ij}_{\scr (1)}
	=
	\delta\pi^{ij} - \frac{1}{3} \delta^{ij} \delta \pi
	\, .
\label{LinearMinkowskiSecondaryTraceless}
\end{equation}
As anticipated from the fact that~$\overline{\rho}\!=\!0$ for Minkowski space,
and from the general results for the Poisson brackets between traceless constraints
in~(\ref{TracelessPoisson}), the two linearised traceless constraints 
are found to commute,
\begin{equation}
\bigl\{ \Phi^{ij}_{\scr (1)} [f_{ij}] , \Psi^{kl}_{\scr (1)} [s_{kl}] \bigr\}
	\approx 0 \, .
\label{MinComm}
\end{equation}
This makes all the constraints first-class, and the conservation of the secondary constraint
generates no further constraints.

Naively speaking, we would now count~$N_{\rm 1st}\!=\!14$ first-class constraints
and~$N_{\rm 2nd}\!=\!0$ second-class ones. Given that the number of canonical variables 
in our Hamiltonian formulation is~$N_{\rm can}\!=\!24$, this would produce a paradoxical 
answer of~$N_{\rm phy}\!=\!-2$ propagating degrees of freedom. Of course, this counting 
is not correct, and the correct counting is a little more subtle.
We should notice that the linearised momentum constraint 
in~(\ref{LinearMinkConstraints}) also changes character compared to its counterpart
in the full theory. Because of the 
secondary traceless constraint~(\ref{LinearMinkowskiSecondaryTraceless}),
it is expressible as a gradient of a scalar function,
\begin{equation}
\mathcal{H}_{\scr (1)}^i \longrightarrow - \frac{2}{3} \partial^i \delta \pi \, ,
\end{equation}
which means it loses its transverse part. This implies that the momentum constraint  counts as a single 
first-class constraint instead of three.\footnote{Strictly
speaking the zero mode remains undetermined by the longitudinal momentum constraint,
but here we consider only normalizable modes.} 
This brings the
total number of first-class constraints to~$N_{\rm 1st} \!=\! 12$. 
Therefore, for the action truncated at quadratic order,
according to formula in~(\ref{Nphy}), one counts no propagating degrees of freedom,
\begin{equation}
N_{\rm phy} = \frac{1}{2} \Bigl( 24 - 2 \!\times\! 12 - 0 \Bigr) = 0 \, .
\end{equation}
Thus, we have uncovered the mechanism behind emptying the spectrum of linearised
perturbations around Minkowski space in the theory that propagates three degrees of
freedom otherwise. While truncating the action at higher orders than quadratic will 
remove this artifact, insisting on perturbation theory around Minkowski space will not 
uncover any degrees of freedom at any order, as we show in the following subsection.

%%%%%%%%%%%%%%%%%%%%%%%%%%%%%%%%%%%%%%%%%%%%%%%
%%%%%%%%%%%%%%%%%%%%%%%%%%%%%%%%%%%%%%%%%%%%%%%
%%%	NONLINEAR PERTURBATIONS
%%%%%%%%%%%%%%%%%%%%%%%%%%%%%%%%%%%%%%%%%%%%%%%
%%%%%%%%%%%%%%%%%%%%%%%%%%%%%%%%%%%%%%%%%%%%%%%
\subsection{Nonlinear perturbations}
\label{subsec: Nonlinear perturbations}

There is a difference between (i) truncating the action at some nonlinear order
and then solving the equations of motion exactly, and (ii) solving the equations of motion
perturbatively as a power series in small fluctuations. Provided that the truncation does
not interfere with constraints, the truncated theory should remain faithful to the full
one when it comes to counting the degrees of freedom. However, this approach is
not directly applicable here because the truncation in the powers of fluctuations preserves
diffeomorphisms only perturbatively. The latter strategy, that we consider here, is not
guaranteed to remain faithful, and in fact, as we shall see, consistently yields no degrees 
of freedom at an arbitrary order in perturbations. This result contradicts the
results of the full Hamiltonian analysis given in Sec.~\ref{Hamiltonian analysis}, 
that applies arbitrarily close to 
Minkowski space. Rather than being a physical property of the theory in the vicinity of 
Minkowski space, we should conclude that perturbation theory centered around Minkowski 
space is not the appropriate scheme to describe this regime of the theory. The behaviour of 
perturbations in this regime is essentially nonperturbative, as already pointed out 
in~\cite{Karananas:2024hoh,Hell:2025lbl}.

The natural assumption when considering small perturbations around a particular 
background is to quantify their evolution as a power series organized in the powers 
of perturbation fields. If we append a bookkeeping parameter~$\varepsilon$ to each 
perturbation field in~(\ref{BackgroundSplit}), then this means we are looking for 
solutions in the following power series form:
\begin{subequations}
\begin{align}
h_{ij} \approx{}&
	\delta_{ij} 
	+ \varepsilon \delta \overline{h}{}^1_{ij} 
	+ \varepsilon^2 \delta \overline{h}{}^2_{ij} 
	+ \varepsilon^3 \delta \overline{h}{}^3_{ij} 
	+ \dots
	\, ,
\\
\pi^{ij} \approx{}& 
	0
	+ \varepsilon \delta \pi^{ij} 
	+ \varepsilon \delta \overline{\pi}{}_1^{ij} 
	+ \varepsilon^2 \delta \overline{\pi}{}_2^{ij} 
	+ \varepsilon^3 \delta \overline{\pi}{}_3^{ij} 
	+ \dots
	\, ,
\\
\mathcal{K}_{ij} \approx{}&
	0
	+ \varepsilon  \delta \mathcal{K}_{ij} 
	+ \varepsilon \delta \overline{\mathcal{K}}{}^1_{ij} 
	+ \varepsilon^2 \delta \overline{\mathcal{K}}{}^2_{ij} 
	+ \varepsilon^3 \delta \overline{\mathcal{K}}{}^3_{ij} 
	+ \dots
	\, ,
\\
\rho^{ij} \approx{}& 
	0
	+ \varepsilon \delta \rho^{ij}
	+ \varepsilon \delta \overline{\rho}{}_1^{ij} 
	+ \varepsilon^2 \delta \overline{\rho}{}_2^{ij} 
	+ \varepsilon^3 \delta \overline{\rho}{}_3^{ij} 
	+ \dots
	\, ,
\\
N \approx{}& 
	1 
	+ \varepsilon  \delta \overline{N}_1
	+ \varepsilon^2  \delta \overline{N}_2 
	+ \varepsilon^3  \delta \overline{N}_3
	+ \dots
	\, ,
\\
N_i \approx{}&
	0
	+ \varepsilon  \delta \overline{N}{}_1^i
	+ \varepsilon^2  \delta \overline{N}{}_2^i
	+ \varepsilon^3  \delta \overline{N}{}_3^i
	+ \dots
	\, ,
\\
\lambda_{ij} \approx{}& 
	0
	+ \varepsilon \delta \overline{\lambda}{}^1_{ij} 
	+ \varepsilon^2 \delta \overline{\lambda}{}^2_{ij} 
	+ \varepsilon^3 \delta \overline{\lambda}{}^3_{ij} 
	+ \dots
	\, ,
\end{align}
\end{subequations}
where there is no~$\varepsilon$ dependence except for the explicitly indicated.
Organizing the theory in powers of~$\varepsilon$ we can derive the action, and 
consequently dynamical equations, for each order of the perturbative correction.

In Sec.~\ref{subsec: Linear perturbations} we found the solutions for the first order:
there are no propagating degrees of freedom, two canonical momenta are found to vanish,
\begin{equation}
\delta \overline{\pi}{}_1^{ij} = 0 \, ,
\qquad \quad
\delta \overline{\rho}{}_1^{ij} = 0 \, ,
\end{equation}
and all the other first order perturbations are undetermined, and depend on the gauge 
choice. This, together with the background fields, serves as a starting point for determining
the properties of higher order perturbations.

\paragraph{Second order perturbation.}
We proceed to find the second order perturbation by shifting the fields in the original action
by the background plus first order solution,
\begin{align}
&
h_{ij} \longrightarrow
	\delta_{ij}
	+ \varepsilon \delta \overline{h}{}^1_{ij}
	+ \varepsilon^2 \delta h_{ij}
	\, ,
\quad\ \
\pi^{ij}\longrightarrow
	\varepsilon^2 \delta \pi^{ij} \, ,
\quad\ \
\mathcal{K}_{ij} \longrightarrow
	\varepsilon \delta \overline{\mathcal{K}}{}^1_{ij}
	+ \varepsilon^2 \delta \mathcal{K}_{ij}
	\, ,
\quad\ \
\rho^{ij} \longrightarrow \varepsilon^2  \delta \rho^{ij} \, ,
\nonumber\\
&
N \longrightarrow
	1
	+ \varepsilon \delta \overline{N}{}_1
	+ \varepsilon^2 \delta N
	\, ,
\qquad
N_i \longrightarrow
	\varepsilon \delta \overline{N}{}^1_i
	+ \varepsilon^2 \delta N_i
	\, ,
\qquad
\lambda_{ij} \longrightarrow
	\varepsilon \delta \overline{\lambda}{}^1_{ij}
	+ \varepsilon^2 \delta \lambda_{ij}
	\, .
\label{MinkShift2}
\end{align}
Solving for the~$\varepsilon$-independent part of the new dynamical variables will then
determine the second order perturbation. The equations of motion at this order are encoded
in the lowest order shifted action, that is obtained by plugging in~(\ref{MinkShift2})
into~(\ref{CanonicalAction}) and keeping only relevant terms,
\begin{align}
\MoveEqLeft[3]
\mathscr{S}_2
	\bigl[ \delta N, \delta N_i, \delta \lambda_{ij}, \delta h_{ij}, \delta \pi^{ij}, 
		\delta \mathcal{K}_{ij}, \delta \rho^{ij} \bigr] 
\nonumber \\
={}&
	\varepsilon^4 \! \int\! d^4x \, \Bigl[
		\delta \pi^{ij} \delta \dot{h}{}_{ij}
		+
		\delta \rho^{ij} \delta \dot{\mathcal{K}}{}_{ij}
		-
		\mathscr{H}_2
		-
		\delta N \mathcal{H}_{\scr (1)}
		-
		\delta N_i \mathcal{H}^i_{\scr (1)}
		-
		\delta \lambda_{ij} \Phi^{ij}_{\scr (1)}
		\Bigr]
		+
		\mathcal{O}(\varepsilon^5)
		\, .
\end{align}
We see that the dynamics of second order perturbations is given by the quadratic
action. The primary constraints are linear, the same ones as for linear perturbations
in~(\ref{LinearMinkConstraints}). The quadratic part of the Hamiltonian 
density,~$\mathscr{H}_2 \!=\! \mathcal{H}_{\scr (2)} \!+\! \Delta\mathscr{H}_2$,
is also the same as for first order perturbations in~(\ref{Mink1stH}), 
but in addition a linear part appears,
\begin{align}
\Delta \mathscr{H}_2
	={}&
\frac{1}{6}
	\Bigl[
	\bigl( \delta \overline{\mathcal{K}}{}_1 \bigr)^2
	- 3 \delta \overline{\mathcal{K}}{}_1^{ij} \delta \overline{\mathcal{K}}{}^1_{ij}
	-
	2 \partial_i \bigl( \delta \overline{h}{}_1^{ij} \partial^k \delta \overline{h}{}^1_{jk} \bigr)
	+
	\partial_i \bigl( \delta \overline{h}{}_1^{ij} \partial_j \delta \overline{h}_1 \bigr)
	+
	\partial^k \bigl( \delta \overline{h}{}_1^{ij} \partial_k \delta \overline{h}{}^1_{ij} \bigr)
\nonumber \\
&
	+
	\bigl( \partial_i \delta \overline{h}{}_1^{ik} \bigr)
		\bigl( \partial^j \delta \overline{h}{}^1_{jk} \bigr)
	- \frac{1}{2}
		\bigl( \partial^i \delta \overline{h}{}_1^{jk} \bigr) \bigl( \partial_j \delta \overline{h}{}^1_{ik} \bigr)
	- \frac{1}{4} \bigl( \partial^i \delta \overline{h}_1 \bigr) \bigl( \partial_i \delta \overline{h}_1 \bigr)
	-
	\frac{1}{4}
		\bigl( \partial^k \delta \overline{h}{}_1^{ij} \bigr)
		\bigl( \partial_k \delta \overline{h}{}^1_{ij} \bigr)
\nonumber \\
&
	+
	\frac{1}{3} \bigl( 3 \delta \overline{N}_1 + \delta \overline{h}_1 \bigr)
		\bigl( \partial^i \partial^j \delta \overline{h}{}^1_{ij}
			- \partial^k \partial_k \delta \overline{h}_1
			\bigr)
	+
	\bigl( 2 \partial^i \delta \overline{h}{}^1_{ik}
			- \partial_k \delta \overline{h}_1 \bigr)
		\partial^k \delta \overline{N }_1
\nonumber \\
&
	+
	2
	\bigl(
	\delta \overline{h}{}^1_{ij} \partial^i \partial^j \delta \overline{N }_1
	-
	\frac{1}{3}
	\delta \overline{h}_1 \partial^i \partial_i \delta \overline{N }_1
	\bigr)
	+
	2 \delta \overline{N}{}_i^1
		\partial^i \delta \overline{\mathcal{K}}_1
	+
	4 \delta \overline{\mathcal{K}}{}_1^{ij}
		\partial_i \delta \overline{N}{}^1_j
	\Bigr]
	\delta\rho
\nonumber \\
&
	-
	\delta \overline{N}{}^k_1
		\bigl( 2\partial_i \delta \overline{h}{}^1_{jk}
			- \partial_k \delta \overline{h}{}^1_{ij} \bigr)
		\delta\pi^{ij}
	\, .
\label{AdditionalLinear1}
\end{align}
Apart from this linear part in the Hamiltonian density, the canonical action for second  order
perturbations is exactly the same as for the first order perturbations~(\ref{FirstOrderAction}). 

While the linear part~(\ref{AdditionalLinear1}) influences the equations of motion
for~$\delta h_{ij}$ and~$\delta \mathcal{K}_{ij}$, it does not change the structure of constraints. Namely, the
conservation of primary constraints generates the same secondary traceless 
constraint~(\ref{LinearMinkowskiSecondaryTraceless}) as found at linear order.
The additional linear piece~(\ref{AdditionalLinear1}) does not influence the dynamics
of the secondary constraint either, and its conservation generates no further
constraints. Thus, we end up with the same first-class constraints as for linear order,
where the momentum constraint loses its transverse part in the same manner.
Therefore, at second order in perturbations we again get that the two canonical 
momenta vanish,
\begin{equation}
\delta \overline{\rho}{}_2^{ij} = 0 \, ,
\qquad\quad
\delta \overline{\pi}{}_2^{ij} = 0 \, ,
\end{equation}
and all the other dynamical variables remain undetermined and dependent on the gauge.

\paragraph{Third order perturbation.}
The third order perturbation is found by first shifting the fields by the background 
solution and solutions for the first two orders of perturbations,
\begin{align}
&
h_{ij} \longrightarrow
	\delta_{ij}
	+ \varepsilon \delta \overline{h}{}^1_{ij}
	+ \varepsilon^2 \delta \overline{h}{}^2_{ij}
	+ \varepsilon^3 \delta h_{ij}
	\, ,
\qquad
\pi^{ij}\longrightarrow
	\varepsilon^3 \delta \pi^{ij} \, ,
\qquad
\mathcal{K}_{ij} \longrightarrow
	\varepsilon \delta \overline{\mathcal{K}}{}^1_{ij}
	+ \varepsilon^2 \delta \overline{\mathcal{K}}{}^2_{ij}
	+ \varepsilon^3 \delta \mathcal{K}_{ij}
	\, ,
\nonumber \\
&
\rho^{ij} \longrightarrow \varepsilon^3  \delta \rho^{ij} \, ,
\qquad
N \longrightarrow
	1
	+ \varepsilon \delta \overline{N}{}_1
	+ \varepsilon^2 \delta \overline{N}{}_2
	+ \varepsilon^3 \delta N
	\, ,
\qquad
N_i \longrightarrow
	\varepsilon \delta \overline{N}{}^1_i
	+ \varepsilon^2 \delta \overline{N}{}^2_i
	+ \varepsilon^3 \delta N_i
	\, ,
\nonumber \\
&
\lambda_{ij} \longrightarrow
	\varepsilon \delta \overline{\lambda}{}^1_{ij}
	+ \varepsilon^2 \delta \overline{\lambda}{}^2_{ij}
	+ \varepsilon^3 \delta \lambda_{ij}
	\, .
\label{MinkShift3}
\end{align}
Plugging these shifted fields  into the action produces the action for cubic perturbations,
\begin{align}
\MoveEqLeft[3]
\mathscr{S}_3
	\bigl[ \delta N, \delta N_i, \delta \lambda_{ij}, \delta h_{ij}, \delta \pi^{ij}, 
		\delta \mathcal{K}_{ij}, \delta \rho^{ij} \bigr] 
\nonumber \\
={}&
	\varepsilon^6 \! \int\! d^4x \, \Bigl[
		\delta \pi^{ij} \delta \dot{h}{}_{ij}
		+
		\delta \rho^{ij} \delta \dot{\mathcal{K}}{}_{ij}
		-
		\mathscr{H}_3
		-
		\delta N \mathcal{H}_{\scr (1)}
		-
		\delta N_i \mathcal{H}^i_{\scr (1)}
		-
		\delta \lambda_{ij} \Phi^{ij}_{\scr (1)}
		\Bigr]
		+
		\mathcal{O}(\varepsilon^7)
		\, .
\end{align}
The primary constraints are once more unchanged, and so is the quadratic part of the 
Hamiltonian,~$\mathscr{H}_3 \!=\! \mathcal{H}_{\scr(2)} \!+\! \Delta \mathscr{H}_3$. 
The only updated part compared to the previous lower order is the
linear part of the Hamiltonian density, that takes the same form as at the quadratic level,
\begin{equation}
\Delta \mathscr{H}_3
	=
	A_3 \delta\rho + B^3_{ij}\delta \pi^{ij}
	\, ,
\end{equation}
only with updated coefficients~$A_3$ and~$B^3_{ij}$ that depend on the solutions of 
lower orders. The concrete expressions for these coefficients are immaterial for 
for the dynamics of constraints. We obtain the same constraint structure as for the linear 
order, implying again that momenta vanish,
\begin{equation}
\delta \overline{\rho}{}_2^{ij} = 0 \, ,
\qquad\quad
\delta \overline{\pi}{}_2^{ij} = 0 \, ,
\end{equation}
and that other dynamical variables are undetermined.

\paragraph{\boldmath $n$-th order perturbation.}
The pattern of behaviour observed for low perturbative orders continues to higher orders.
It is proven in Appendix~\ref{sec: Higher order perturbations} that the quadratic part of the action
for the~$n$-th order perturbation has the same form at every order, just as at first 
order. The part of the action that is updated at each successive order is only
the linear part, which here contributes to the Hamiltonian density only.

Here in addition we observe another pattern, which is the vanishing of the perturbation 
of the canonical momenta~$\rho^{ij}$ and~$\pi^{ij}$ at each order.
It then follows that, after plugging in the variables shifted by the solutions for
first~$(n\!-\!1)$ orders, the action for the~$n$-th order perturbation reads
\begin{align}
\MoveEqLeft[3]
\mathscr{S}^n
	\bigl[ \delta N, \delta N_i, \delta \lambda_{ij}, \delta h_{ij}, \delta \pi^{ij}, 
		\delta \mathcal{K}_{ij}, \delta \rho^{ij} \bigr] 
\nonumber \\
={}&
	\varepsilon^{2n} \! \int\! d^4x \, \Bigl[
		\delta \pi^{ij} \delta \dot{h}{}_{ij}
		+
		\delta \rho^{ij} \delta \dot{\mathcal{K}}{}_{ij}
		-
		\mathscr{H}_n
		-
		\delta N \mathcal{H}_{\scr (1)}
		-
		\delta N_i \mathcal{H}^i_{\scr (1)}
		-
		\delta \lambda_{ij} \Phi^{ij}_{\scr (1)}
		\Bigr]
		+
		\mathcal{O}(\varepsilon^{2n+1})
		\, .
\end{align}
The Hamiltonian density always has the same quadratic 
part,~$\mathscr{H}_n \!=\! \mathcal{H}_{\scr (2)} \!+\! \Delta \mathscr{H}_n$,
but the linear part changes from order to order,
\begin{equation}
\mathscr{H}_n = A_n \delta\rho + B^n_{ij} \delta \pi^{ij} \, ,
\end{equation}
where its form remains the same, and just the coefficients~$A_n$ and~$B_{ij}^n$
get updated. It is straightforward to prove by induction that 
this form of the Hamiltonian density is true at every order.
It is then easy to see that the linear part of the Hamiltonian density does not participate
in determining the conservation of constraints. That is why the conservation of primary
constraints at~$n$-th order generate a secondary constraint~$\Psi_{\scr (1)}^{ij}$,
such that all constraints are first-class, and that the momentum constraint becomes 
longitudinal.

\paragraph{Implications and interpretation.}
The perturbative expansion of small perturbations around Minkowski space 
consistently produces the same constraint structure at each order,
and yields an empty spectrum of propagating degrees of freedom. It might
seem that this implies a discontinuous jump in the number of degrees of freedom
when we approach close to Minkowski space.
However, this conclusion would contradict the general result of the constraint analysis
in Sec.~\ref{Hamiltonian analysis} that is valid arbitrarily close to Minkowski space.
The resolution of this seeming paradox is found upon closer examination of the perturbative
analysis. The relevant fact is the vanishing of~$\rho$ to all orders, which is equivalent
to the vanishing Ricci scalar at all orders. What this means is that perturbation theory does
not admit small variations of the Ricci scalar, and forces us to exact Minkowski background 
at each order. This means
that expansion in small perturbations is not appropriate to probe the vicinity of 
Minkowski space, where the dynamics of propagating degrees of freedom becomes 
nonperturbative (i.e.~perturbations become strongly coupled), 
and is missed by the perturbative expansion centered on Minkowski background. 
Rather, more sophisticated analytic methods are called for to analyse this regime.

%%%%%%%%%%%%%%%%%%%%%%%%%%%%%%%%%%%%%%%%%%%%%%%
%%%%%%%%%%%%%%%%%%%%%%%%%%%%%%%%%%%%%%%%%%%%%%%
%%%%%%%%%%%%%%%%%%%%%%%%%%%%%%%%%%%%%%%%%%%%%%%
%%%	PERTURBATIONS AROUND OTHER SINGULAR BACKGROUNDS
%%%%%%%%%%%%%%%%%%%%%%%%%%%%%%%%%%%%%%%%%%%%%%%
%%%%%%%%%%%%%%%%%%%%%%%%%%%%%%%%%%%%%%%%%%%%%%%
%%%%%%%%%%%%%%%%%%%%%%%%%%%%%%%%%%%%%%%%%%%%%%%
\section{Perturbations around other singular backgrounds}
\label{sec: Perturbations around other singular backgrounds}

The results of the Hamiltonian analysis of Sec.~\ref{Hamiltonian analysis} suggests 
that Minkowski spacetime is not an isolated background exhibiting a strong coupling feature
for linearised perturbations. Rather, we expect those to be all backgrounds for which
the bracket~(\ref{TracelessPoisson}) vanishes, signaling the change of second-class
constraints to first-class. In this section we first provide a couple of examples of such 
spacetimes, and then show that all traceless-Ricci backgrounds, i.e.~ backgrounds with
a vanishing Ricci scalar, exhibit the same strong coupling feature.

%%%%%%%%%%%%%%%%%%%%%%%%%%%%%%%%%%%%%%%%%%%%%%%
%%%%%%%%%%%%%%%%%%%%%%%%%%%%%%%%%%%%%%%%%%%%%%%
%%%	SCHWARZSCHILD SPACETIME
%%%%%%%%%%%%%%%%%%%%%%%%%%%%%%%%%%%%%%%%%%%%%%%
%%%%%%%%%%%%%%%%%%%%%%%%%%%%%%%%%%%%%%%%%%%%%%%
\subsection{Schwarzschild spacetime}
\label{sec: Schwarzschild spacetime}

Spherically symmetric and static black hole spacetime is a vacuum solution of Einstein's 
general relativity, and is thus a Ricci-flat spacetime. It is consequently also
a solution of the pure~$R^2$ theory, where, as we show, it possesses an empty
spectrum of linearised perturbations. Schwarzschild spacetime is described by the line 
element
\begin{equation}
ds^2 = \overline{g}_{\mu\nu} dx^\mu dx^\nu
	=
	- 
	f(r) dt^2
	+
	\frac{dr^2}{ f(r) }
	+
	r^2 \Bigl( d\theta^2 + \sin^2(\theta) d\varphi^2 \Bigr)
	\, ,
	\qquad
	f(r) = 1 - \frac{r_{\scr \rm S}}{r}
	\, ,
\end{equation}
where~$r_{\scr \rm S} \!=\! 2 G_{\scr \rm N} M$ is the Schwarzschild radius,~$M$ is the black hole 
mass, and~$G_{\scr \rm N}$ is the Newton constant. The lapse and shift variables of the ADM 
decomposition for the Schwarzschild metric are read off from this diagonal
line element as~$\overline{N} \!=\! \sqrt{ f(r) }$ 
and~$\overline{N}_i = 0$, and the induced spatial metric is diagonal and time-independent, 
as inferred from the line element on equal time hypersurfaces,
\begin{equation}
d\ell^2 
	= 
	\overline{h}_{ij} dx^i dx^j
	=
	\frac{dr^2}{f(r)}
	+
	r^2 \Bigl( d\theta^2 + \sin^2(\theta) d\varphi^2 \Bigr)
	\, .
\end{equation}
It follows that the extrinsic curvature vanishes,~$\overline{\mathcal{K}}_{ij} \!=\! 0$, 
and consequently from Eq.~(\ref{EOM3}) that~$\overline{\lambda}_{ij}\!=\!0$. 
Then it follows from vanishing of the Ricci scalar~$\overline{R}\!=\!0$ from 
Eqs.~(\ref{Fdef}) and~(\ref{Rdecomposition}) that
\begin{equation}
\overline{\mathcal{R}}_{ij}
=
\frac{2}{\overline{N}} \overline{\nabla}_i \overline{\nabla}_j \overline{N}
\, .
\end{equation}
The trace of Eq.~(\ref{EOM2}), together with the primary traceless constraint, then
tells us that~$\overline{\rho}{}^{ij}\!=\!0$, and subsequently Eq.~(\ref{EOM1}) 
implies~$\overline{\pi}{}^{ij} \!=\! 0$. 

The canonical action for linear perturbations around Schwarzschild spacetime reads
\begin{align}
\MoveEqLeft[5]
\mathscr{S}_{\scr (2)}
	\bigl[ \delta N, \delta N_i, \delta \lambda_{ij}, \delta h_{ij}, \delta \pi^{ij}, 
		\delta \mathcal{K}_{ij}, \delta \rho^{ij} \bigr] 
\nonumber \\
={}&
	\int\! d^4x \, \Bigl[
		\delta \pi^{ij} \delta \dot{g}_{ij}
		+
		\delta \rho^{ij} \delta \dot{\mathcal{K}}_{ij}
		-
		\mathscr{H}_{\scr (2)}
		-
		\delta N \mathcal{H}_{\scr (1)}
		-
		\delta N_i \mathcal{H}^i_{\scr (1)}
		-
		\delta \lambda_{ij} \Phi^{ij}_{\scr (1)}
		\Bigr]
		\, ,
\end{align}
where only the second perturbation of the Hamiltonian constraint contributes to the 
quadratic canonical Hamiltonian density,
\begin{equation}
\mathscr{H}_{\scr (2)} 
	=
	\overline{N}\sqrt{\overline{h}}\biggl[
	\frac{1}{144} \Bigl( \frac{\delta\rho}{ \sqrt{\overline{h}} } \Bigr)^{\!2}
	-
	2 \delta \mathcal{K}_{ij} \frac{ \delta\pi^{ij} }{ \sqrt{\overline{h}} }
	+
	\frac{1}{6}
	\bigl( \overline{\nabla}_i \delta h \bigr)
	\overline{\nabla}{}^i \Bigl( \frac{ \delta \rho }{ \sqrt{h} } \Bigr)
	+
	\frac{1}{6}
	\delta h_{ij}
	\Bigl( \overline{\nabla}{}^i \overline{\nabla}{}^j - \frac{1}{2} \overline{\mathcal{R}}{}^{ij} \Bigr)
		\frac{ \delta \rho }{ \sqrt{\overline{h}} }
	\biggr]
	\, ,
\end{equation}
where~$\delta h \!=\! \overline{h}{}^{ij} \delta h_{ij}$, and 
where~$\overline{N}\sqrt{\overline{h}} \!=\! r^2 \sin(\theta)$
is the volume Jacobian of the spherical coordinate system.
The linearised primary constraints are given by
\begin{equation}
\mathcal{H}_{\scr (1)} =
	 -
	\frac{1}{3} \sqrt{\overline{h} } \,
	\Bigl( \overline{\nabla}{}^i \overline{\nabla}_i
		- \frac{1}{2} \overline{\mathcal{R}} \Bigr)
		\frac{ \delta\rho }{ \sqrt{ \overline{h} } }
	\, ,
\quad \ \
\mathcal{H}^i_{\scr (1)} =
	- 2
	\sqrt{\overline{h}}\,
	 \overline{\nabla}_j \Bigl( \frac{ \delta\pi^{ij} }{ \sqrt{ \overline{h} } } \Bigr)
	\, ,
\quad \ \
\Phi^{ij}_{\scr (1)} = \delta \rho^{ij} - \frac{1}{3} \overline{h}^{ij} \delta \rho \, .
\label{SchPrimary}
\end{equation}
All primary constraints are mutually first-class, and their conservation,
\begin{subequations}
\begin{align}
\dot{\mathcal{H}}_{\scr (1)} ={}&
	- \frac{2}{3} \sqrt{\overline{h}} \,
	\Bigl[
	\overline{N} \, \overline{\nabla}{}^i \overline{\nabla}_i
		+ 2 \bigl( \overline{\nabla}{}^i \overline{N} \bigr) \overline{\nabla}_i 
	\Bigr]
	\frac{\delta\pi}{\sqrt{\overline{h}}}
	\approx
	2 \sqrt{\overline{h}} \, \Bigl[
		\overline{N} \, \overline{\nabla}_i \overline{\nabla}_j
		+ 2 \bigl( \overline{\nabla}_i \overline{N} \bigr) \overline{\nabla}_j
		\Bigr]
	\frac{\Psi^{ij}_{\scr (1)}}{ \sqrt{ \overline{h}} }
	\, ,
\\
\dot{\mathcal{H}}_{\scr (1)}^i \approx{}& 0 \,,
\qquad \qquad
\dot{\Phi}^{ij}_{\scr (1)} \approx
	2 \overline{N} \Psi^{ij}_{\scr (1)}
\end{align}
\end{subequations}
generates a secondary traceless constraint,
\begin{equation}
\Psi^{ij}_{\scr (1)}
	=
	\delta\pi^{ij} - \frac{1}{3} \overline{h}^{ij} \delta \pi \, .
\end{equation}
The conservation of this secondary constraint generates no further constraints. Moreover,
it is first-class with all the primary constraints, including the traceless one,
\begin{equation}
\bigl\{ \Phi^{ij}_{\scr (1)} [f_{ij}] , \Psi^{kl}_{\scr (1)} [s_{kl}] \bigr\}
	\approx 0 \, .
\end{equation}
This makes all constraints first-class; the naive count gives~$N_{\rm 1st} \!=\! 14$ of them.
However, the secondary constraint removes the transverse part of the
momentum constraint in~(\ref{SchPrimary}),
\begin{equation}
\mathcal{H}^i_{\scr (1)} \approx
	- \frac{2}{3}
	\sqrt{\overline{h}}\,
	 \overline{\nabla}{}^i \Bigl( \frac{ \delta\pi }{ \sqrt{ \overline{h} } } \Bigr)
	 \, ,
\end{equation}
reducing the number of first-class constraints to~$N_{\rm 1st} \!=\! 12$. 
This brings us to conclude that there 
are no linear degrees of freedom around Schwarzschild black hole spacetime, owing to 
the same mechanism observed for linear perturbations around Minkowski space.
This result contradicts previous work~\cite{Dioguardi:2020nxr}
that reported the stability analysis of a linearised scalar degree of freedom in
pure~$R^2$ theory around the Schwarzschild black hole.

%%%%%%%%%%%%%%%%%%%%%%%%%%%%%%%%%%%%%%%%%%%%%%%
%%%%%%%%%%%%%%%%%%%%%%%%%%%%%%%%%%%%%%%%%%%%%%%
%%%	RADIATION-DOMINATED COSMOLOGICAL SPACETIME
%%%%%%%%%%%%%%%%%%%%%%%%%%%%%%%%%%%%%%%%%%%%%%%
%%%%%%%%%%%%%%%%%%%%%%%%%%%%%%%%%%%%%%%%%%%%%%%
\subsection{Radiation-dominated cosmological spacetime}
\label{ref: Radiation-dominated cosmological spacetime}

Homogeneous, isotropic, and spatially flat expanding cosmological spacetime is described
by the Friedmann-Lama\^{i}tre-Robertson-Walker (FLRW) line element,
\begin{equation}
ds^2 = g_{\mu\nu} dx^\mu dx^\nu
	= - dt^2 + a^2(t) d\vec{x}^{\,2} \, ,
\label{FLRWline}
\end{equation}
where the scale factor~$a(t)$ encodes the dynamics of the expansion. It is not a vacuum 
solution of Einstein's general relativity, and in general its Ricci tensor does not vanish.
In the special case when the expansion is sourced by conformal matter, 
the Ricci tensor is traceless. This is the case 
of radiation-dominated universe, where the Hubble rate,~$H=\dot{a}/a$, 
satisfies~$\dot{H}=-2H^2$. Such a spacetime is indeed a vacuum solution of the 
pure~$R^2$ theory, on account of its vanishing Ricci scalar.

The background ADM variables~(\ref{BackgroundSplit}) for this spacetime are
inferred from the line element~(\ref{FLRWline}), the canonical equations of 
motion~(\ref{EOM1})--(\ref{EOM4}), and 
constraints~(\ref{PrimaryConstraints}) and~(\ref{SecondaryTracelessConstraint}),
\begin{equation}
\overline{N}=1 \, ,
\quad\
\overline{N}_i=0 \, ,
\quad\
\overline{h}_{ij} = a^2 \delta_{ij} \, ,
\quad\
\overline{\pi}{}^{ij} =0 \, ,
\quad\
\overline{\mathcal{K}}_{ij} = - H \overline{h}_{ij}  \, ,
\quad\
\overline{\rho}{}^{ij} = 0 \, ,
\quad\
\overline{\lambda}_{ij} = 0 \, .
\end{equation}
The canonical action for linearised perturbations around this background is then given by
\begin{align}
\MoveEqLeft[5]
\mathscr{S}_{\scr (2)}
	\bigl[ \delta N, \delta N_i, \delta \lambda_{ij}, \delta h_{ij}, \delta \pi^{ij}, 
		\delta \mathcal{K}_{ij}, \delta \rho^{ij} \bigr] 
\nonumber \\
={}&
	\int\! d^4x \, \Bigl[
		\delta \pi^{ij} \delta \dot{h}_{ij}
		+
		\delta \rho^{ij} \delta \dot{\mathcal{K}}_{ij}
		-
		\mathscr{H}_{\scr(2)}
		-
		\delta N \mathcal{H}_{\scr (1)}
		-
		\delta N_i \mathcal{H}^i_{\scr (1)}
		-
		\delta \lambda_{ij} \Phi^{ij}_{\scr (1)}
		\Bigr]
		\, .
\end{align}
where the linearised constraints are given by
\begin{equation}
\mathcal{H}_{\scr (1)}
	=
	-
	\frac{1}{3} \partial^i \partial_i \delta\rho
	+
	2 H a^2 \delta \pi
	\, ,
\qquad
\mathcal{H}^i_{\scr (1)}
	=
	- 2 \partial_j \delta\pi^{ij}
	+
	\frac{2}{3} H \partial^i \delta\rho
	\, ,
\qquad
\Phi^{ij}_{\scr (1)}
	=
	\delta \rho^{ij} - \frac{1}{3} \delta^{ij} \delta \rho \, .
\end{equation}
Note that in this subsection the indices are raised and lowered by the Kronecker delta symbol,
and traces are defined accordingly,~$\delta \pi \!=\! \delta_{ij} \delta \pi^{ij}$,~$\delta\rho \!=\!\delta_{ij} \delta \rho^{ij}$. The quadratic Hamiltonian density receives contribution
from the second perturbation of the Hamiltonian constraint only,
\begin{equation}
\mathscr{H}_{\scr(2)}
	=
	\frac{a}{144} \delta\rho^2
	-
	2 \delta\mathcal{K}_{ij} \delta\pi^{ij}
	+
	\frac{1}{6a^2} \delta\rho
		\bigl( \partial^i \partial^j 
			- \delta^{ij} \partial^k \partial_k \bigr) \delta h_{ij}
	\, ,
\end{equation}
where we have discarded total derivatives.

All primary constraints are mutually first-class. Their conservation,
\begin{equation}
\dot{\mathcal{H}}_{\scr (1)}
	\approx
	- \frac{2}{3} \partial^i \partial_i 
	\bigl( \delta\pi - H \delta \rho \bigr)
	\approx
	2 \partial_i \partial_j \Psi_{\scr (1)}^{ij}
	\, ,
\qquad
\dot{\mathcal{H}}^i_{\scr (1)}
	\approx
	- 4 H \partial_j \Psi_{\scr (1)}^{ij}
	\, ,
\qquad
\dot{\Phi}^{ij}_{\scr (1)}
	\approx
	2 \Psi_{\scr (1)}^{ij}
	\, ,
\end{equation}
generates a secondary traceless constraint,
\begin{equation}
\Psi^{ij}_{\scr (1)} = \delta \pi^{ij} - \frac{1}{3} \delta^{ij} \delta \pi 
	\, .
\end{equation}
This constraint is also first-class with all others, and its conservation generates no further
constraints. Furthermore, this secondary constraint makes the momentum constraint 
longitudinal,
\begin{equation}
\mathcal{H}^i_{\scr (1)}
	\approx
	- \frac{2}{3} \partial^i \bigl( \delta\pi - H \delta\rho \bigr)
	\, ,
\end{equation}
meaning it should really be interpreted as a scalar 
constraint~$\delta\pi \!\approx\! H \delta \rho$, which then turns the Hamiltonian 
constraint into
\begin{equation}
\mathcal{H}_{\scr (1)} \approx - \frac{1}{3} \bigl( \partial^i \partial_i - 6a^2 H^2 \bigr) \delta\rho
	\approx 0 \, .
\end{equation}
This constraint takes the form of the modified Helmholtz equation with a time-dependent 
mass, and is equivalent to~$\delta\rho\!\approx\!0$, which then 
implies~$\delta\pi\!\approx\!0$. The feature of losing two constraints from the 
momentum constraint brings  the total number of first-class constraints 
to~$N_{\rm 1st}\!=\!12$, which implies no propagating degrees of freedom in the linearised 
spectrum.

%%%%%%%%%%%%%%%%%%%%%%%%%%%%%%%%%%%%%%%%%%%%%%%
%%%%%%%%%%%%%%%%%%%%%%%%%%%%%%%%%%%%%%%%%%%%%%%
%%%	GENERAL TRACELESS-RICCI TENSOR SPACETIMES
%%%%%%%%%%%%%%%%%%%%%%%%%%%%%%%%%%%%%%%%%%%%%%%
%%%%%%%%%%%%%%%%%%%%%%%%%%%%%%%%%%%%%%%%%%%%%%%
\subsection{General traceless-Ricci spacetimes}
\label{subsec: General traceless-Ricci spacetimes}

The two examples of preceding subsections support what is
already clear from the bracket~(\ref{TracelessPoisson}) between primary and secondary traceless
constraints: singular behaviour is not innate to perturbations around Minkowski space 
background, but is expected whenever~$\overline{\rho}$ vanishes. 
Together with the constraint in~(\ref{TracelessConstraint}) that is always valid, which implies singular points
will have a vanishing canonical momentum associated to extrinsic curvature,
\begin{equation}
\overline{\rho}{}^{ij} = 0 \, .
\label{VanishingRho}
\end{equation}
By taking the trace of Eq.~(\ref{EOM3}) and inspecting Eqs.~(\ref{Fdef})
and~(\ref{Rdecomposition}),
\begin{equation}
\frac{\rho}{\sqrt{h}} \approx  - 12\Bigl( 2F + K^2 - K_{ij} K^{ij} + \mathcal{R} \Bigr) = -12 R \, ,
\end{equation}
it is possible to interpret such spacetime points as those for which the covariant
Ricci scalar vanishes,~$\overline{R}\!=\!0$. 

While we expect the singular features for 
perturbations to appear locally where this happens, the local characterization of the
constraint structure would require a more detailed analysis. Here we consider general
traceless-Ricci spacetimes, for which~(\ref{VanishingRho}) is true everywhere and for all 
times. For such spacetimes combining the condition in~(\ref{VanishingRho}) with equation 
of motion~(\ref{EOM4}) implies vanishing of the canonical momentum associated to the 
spatial metric,
\begin{equation}
\overline{\pi}{}^{ij} = 0 \, .
\end{equation}
The rest of the variables are not constrained, apart from having to satisfy the equations
of motion that remain from~(\ref{EOM1})--(\ref{EOM2}),
\begin{align}
\dot{\overline{h}}_{ij} \approx{}&
	- 2 \overline{N} \, \overline{\mathcal{K}}_{ij}
	+ 2 \overline{\nabla}_{(i} \overline{N}_{j)}
	\, ,
\\
\dot{\overline{\mathcal{K}}}_{ij} \approx{}&
	\frac{\overline{h}_{ij} }{3}
	\biggl[
	\frac{\overline{N}}{2} \Bigl(
	\overline{\mathcal{K}}{}^2 
		\!-\! 3\overline{\mathcal{K}}{}^{kl} \overline{\mathcal{K}}_{kl} + \overline{\mathcal{R}}
	\Bigr)
	-
	\overline{\nabla}{}^k \overline{\nabla}_k \overline{N}
	+
	\overline{N}{}^k \overline{\nabla}_k \overline{\mathcal{K}}
	+
	2 \overline{\mathcal{K}}{}^{kl} \overline{\nabla}_k \overline{N}_l
	\biggr]
	\!
	+
	\overline{N} \, \overline{\lambda}_{ij}
	\, .
\end{align}

The three spacetimes considered in sections~\ref{subsec: Linear perturbations},
\ref{sec: Schwarzschild spacetime}, and~\ref{ref: Radiation-dominated cosmological spacetime} are just special instances of traceless-Ricci spacetimes, with the first one
being Ricci-flat. Some further 
examples~\cite{Stephani:2003,Griffiths:2009dfa,Muller:2009bw} of exact Ricci-flat spacetimes
are
Kerr spacetime (and Taub-NUT spacetime that generalizes it), %Milne spacetime,
Kasner spacetime, and pp-wave spacetime; all of them are exact vacuum solutions 
of pure~$R^2$ theory and exhibit an empty spectrum of linearised perturbations.

%%%%%%%%%%%%%%%%%%%%%%%%%%%%%%%%%%%%%%%%%%%%%%%
%%%	CANONICAL STRUCTURE OF LINEARISED PERTURBATIONS
%%%%%%%%%%%%%%%%%%%%%%%%%%%%%%%%%%%%%%%%%%%%%%%
\paragraph{Canonical structure of linearised perturbations.}
Shifting the variables as in~(\ref{BackgroundSplit}), and retaining only quadratic terms in the 
canonical action gives
\begin{align}
\MoveEqLeft[7]
\mathscr{S}_{\scr (2)} \bigl[ \delta N, \delta N_i, \delta \lambda_{ij}, \delta h_{ij}, 
	\delta \pi^{ij}, \delta \mathcal{K}_{ij}, \delta \rho^{ij} \bigr] 
\nonumber \\
	={}&
	\int\! d^4x \, \Bigl[
		\delta \pi^{ij} \delta \dot{h}_{ij}
		+
		\delta \rho^{ij} \delta \dot{\mathcal{K}}_{ij}
		-
		\mathscr{H}_{\scr (2)}
		-
		\delta N \mathcal{H}_{\scr (1)}
		-
		\delta N_i \mathcal{H}^i_{\scr (1)}
		-
		\delta \lambda_{ij} \Phi^{ij}_{\scr (1)}
		\Bigr]
		\, ,
\end{align}
where throughout we tacitly shifted and rescaled the perturbation of the Lagrange
multiplier~$\delta \lambda_{ij}$ to absorb all the traceless parts of~$\delta\rho^{ij}$.
The linearised Hamiltonian and momentum constraints are now given by
\begin{align}
\mathcal{H}_{\scr (1)} ={}&
	\sqrt{\overline{h} } \, \biggl[
	-
	2 \overline{\mathcal{K}}_{ij} \frac{ \delta \pi^{ij} }{ \sqrt{ \overline{h} } }
	+
	\frac{ 1 }{6} 
		\Bigl( \overline{\mathcal{K}}^2 
			\!-\! 3\overline{\mathcal{K}}{}^{ij} \overline{\mathcal{K}}_{ij} 
			\!+\! \overline{\mathcal{R}} \Bigr)
		\frac{ \delta\rho }{ \sqrt{ \overline{h} } }
	-
	\frac{1}{3}
	\overline{\nabla}{}^i \overline{\nabla}_i
		\Bigl( \frac{ \delta\rho }{ \sqrt{ \overline{h} } } \Bigr)
	\biggr]
	\, ,
\label{HamiltonianLinear}
\\
\mathcal{H}^i_{\scr (1)} ={}&
	\sqrt{\overline{h}} \, \biggl[
	- 2 \overline{\nabla}_j \Bigl( \frac{ \delta\pi^{ij} }{ \sqrt{ \overline{h} } } \Bigr)
	+
	\frac{1}{3} \frac{ \delta\rho}{ \sqrt{ \overline{h} } } 
		\overline{\nabla}{}^i \overline{\mathcal{K}}
	-
	\frac{2}{3} \overline{\nabla}_j \Bigl( \overline{\mathcal{K}}{}^{ij} 
		\frac{ \delta\rho }{ \sqrt{ \overline{h} }} \Bigr)
	\biggr]
	\, .
\label{MomentumLinear}
\end{align}
and the linearised primary traceless constraint reads
\begin{equation}
\Phi^{ij}_{\scr (1)}
	=
	\delta \rho^{ij} - \frac{1}{3} \overline{h}{}^{ij} \delta \rho \, .
\label{PhiLinear}
\end{equation}
The quadratic canonical Hamiltonian density that governs 
the dynamics of linearized perturbations,
\begin{equation}
\mathscr{H}_{\scr(2)} = \overline{N} \bigl( \mathcal{H}^{\scr (2)} 
	+ \overline{\lambda}_{ij} \Phi^{ij}_{\scr (2)} \bigr)
	+ \overline{N}_i \mathcal{H}^i_{\scr (2)},
\label{secondH}
\end{equation}
is composed of second variations of the primary constraints,
\begin{align}
\mathcal{H}_{\scr (2)}
	={}&
	\sqrt{ \overline{h} } \,
	\biggl[
	\frac{1}{144} \Bigl( \frac{\delta\rho}{ \sqrt{ \overline{h} } } \Bigr)^{\!2}
	-
	2 \delta\mathcal{K}_{ij} \frac{ \delta\pi^{ij} }{ \sqrt{ \overline{h} } }
	+
	\Bigl( \overline{\mathcal{K}}{}^{ij}
			- \frac{1}{3} \overline{h}{}^{ij} \overline{\mathcal{K}} \Bigr)
		\Bigl( \overline{\mathcal{K}}_i{}^k \delta h_{kj}
			- \frac{1}{6} \overline{\mathcal{K}}_{ij} \delta h 
			- \delta \mathcal{K}_{ij}
			 \Bigr)
		\frac{\delta\rho}{ \sqrt{ \overline{h} } }
\nonumber \\
&	\hspace{1cm}
	-
	\frac{1}{6} \Bigl( \overline{\mathcal{R}}{}^{ij} 
			- \frac{1}{3} \overline{h}{}^{ij} \overline{R} \Bigr)
		\delta h_{ij} \frac{ \delta\rho }{ \sqrt{ \overline{h} } }
	+
	\frac{1}{3} \Bigl( \delta_k^{(i} \delta_l^{j)}
			- \frac{1}{3} \overline{h}{}^{ij} \overline{h}_{kl} \Bigr)
		\overline{\nabla}{}^k \Bigl( \delta h_{ij} \overline{\nabla}{}^l
			\frac{\delta\rho}{\sqrt{ \overline{h} }} \Bigr)
\nonumber \\
&	\hspace{1cm}
	+
	\frac{1}{6} \Bigl( \frac{\delta\rho}{\sqrt{ \overline{h} }}
		\overline{\nabla}{}^i \overline{\nabla}{}^j \delta h_{ij}
		+
		\bigl(\overline{\nabla}{}^k \delta h \overline{\nabla}_k \bigr)
			\frac{\delta\rho}{\sqrt{ \overline{h} }}
		\Bigr)
	\biggr]
	\, ,
\label{HamiltonianQuadratic}
\\
\mathcal{H}^i_{\scr(2)}
	={}&
	\sqrt{ \overline{h} } \,
	\biggl[
	-
	2\frac{ \delta \pi^{jk} }{ \sqrt{ \overline{h} } }
	\Bigl( \overline{\nabla}_k \delta h_j{}^i
		- \frac{1}{2} \overline{\nabla}{}^i \delta h_{jk} \Bigr)
	+
	\frac{2}{3}
		\Bigl( \delta^k_j \overline{\nabla}{}^l
			- \frac{1}{3} \overline{h}{}^{kl} \overline{\nabla}_j \Bigr)
	 \Bigl( 
			\overline{\mathcal{K}}{}^{ij} \delta h_{kl}
		\frac{ \delta \rho }{ \sqrt{ \overline{h} } } \Bigr)
\nonumber \\
&	\hspace{1cm}
	-
	\frac{1}{3} \Bigl( \delta h^{ij} - \frac{1}{3} \overline{h}{}^{ij} \delta h \Bigr)
		\bigl( \overline{\nabla}_j \overline{\mathcal{K}} \bigr)
		\frac{\delta\rho}{ \sqrt{ \overline{h} } }
	+
	\frac{1}{3} \frac{\delta\rho}{ \sqrt{ \overline{h} } }
		\Bigl( \overline{\nabla}{}^i \delta\mathcal{K}
			- \delta h_{jk} \overline{\nabla}{}^i \overline{\mathcal{K}}{}^{jk} \Bigr)
\nonumber \\
&	\hspace{1cm}
	+
	\frac{2}{3} \delta h^{ij} \overline{\nabla}{}^k \Bigl(
		\overline{\mathcal{K}}_{jk} \frac{\delta\rho}{ \sqrt{ \overline{h} } }
		\Bigr)
	-
	\frac{2}{3} \overline{\nabla}{}_j
		\Bigl( \delta \mathcal{K}^{ij} \frac{\delta\rho}{ \sqrt{ \overline{h} } } \Bigr)
	\biggr]
	\, ,
\label{MomentumQuadratic}
\\
\Phi^{ij}_{\scr(2)}
	={}&
	\frac{1}{3} \Bigl( \delta h^{ij} 
	- \frac{1}{3} \overline{h}{}^{ij} \delta h \Bigr)
	\delta \rho
	\, .
\label{PhiQuadratic}
\end{align}
where~$\delta h \!=\! \overline{h}{}^{ij} \delta h_{ij}$.

We notice there are no terms quadratic in coordinates in the 
parts~(\ref{HamiltonianQuadratic})--(\ref{PhiQuadratic}) making up the canonical
Hamiltonian. This means that the conservation of 
constraints~(\ref{HamiltonianLinear})--(\ref{PhiLinear}) which are linear
in the perturbed momenta can only give rise to constraints which are also linear in the
perturbed momenta. Indeed, the conservation of linearised primary constraints
\begin{subequations}
\begin{align}
\dot{\mathcal{H}}_{\scr (1)}
	\approx{}&
	2 \sqrt{ \overline{h} } \, \Bigl(
	\overline{N} \, \overline{\nabla}_i 
	+
	2 \overline{\nabla}_i \overline{N} 
	\Bigr)
	\overline{\nabla}_j 
	\Bigl( \frac{ \Psi_{\scr (1)}^{ij} }{ \sqrt{ \overline{h} } } \Bigr)
	+
	2\Bigl( 2\overline{\mathcal{K}}_{ik} \overline{\nabla}_j \overline{N}{}^k
	+
	\overline{N}{}^k \overline{\nabla}_k \overline{\mathcal{K}}_{ij} \Bigr) \Psi_{\scr (1)}^{ij} 
	-
	2 \overline{N} \, \overline{\lambda}_{ij} \Psi_{\scr (1)}^{ij}
	\, ,
\\
\dot{\mathcal{H}}^{i}_{\scr (1)}
	\approx{}&
	4 \sqrt{ \overline{h} } \, \overline{\nabla}_j
		\Bigl( \overline{N} \, \overline{\mathcal{K}}{}^i{}_k
			\frac{\Psi_{\scr (1)}^{jk}}{ \sqrt{ \overline{h} } } \Bigr)
	-
	2 \overline{N} \Psi_{\scr (1)}^{jk} \overline{\nabla}{}^i \overline{\mathcal{K}}_{jk}
	\, ,
\qquad \qquad
\dot{\Phi}^{ij}_{\scr (1)}
	\approx
	2 \overline{N} \Psi^{ij}_{\scr (1)}
	\, ,
\end{align}
\label{GeneralConservation}%
\end{subequations}
implies that
\begin{equation}
\Psi^{ij}_{\scr (1)}
	=
	\delta \pi^{ij} - \frac{1}{3} \overline{h}{}^{ij} \delta \pi 
	+
	\frac{\delta\rho}{6} \Bigl( \overline{\mathcal{K}}{}^{ij} 
		- \frac{1}{3} \overline{h}{}^{ij} \overline{\mathcal{K}} \Bigr)
	\, ,
\end{equation}
appears as a secondary traceless constraint.
Note that in deriving Eqs.~(\ref{GeneralConservation})
it is necessary to use equations of motion~(\ref{EOM1})--(\ref{EOM4}) given that primary 
constraints are now explicitly time-dependent via the background quantities.

All the linearised constraints (being linear in perturbed momenta)
trivially commute, and are therefore all first-class. Consequently, no further constraints are 
generated by the conservation of the secondary constraint.
Naively we would count 14 first-class constraints --- 4 for Hamiltonian and momentum 
constraints, and 10 for primary and secondary traceless constraints. However, that would 
not be consistent, given that the number of canonical variables is 24. It should not be 
overlooked that Hamiltonian and momentum constraints are impacted by the secondary 
traceless constraint, that eliminates the traceless part of~$\delta \pi^{ij}$,
\begin{align}
\mathcal{H}_{\scr (1)} \approx{}&
	\sqrt{\overline{h} } \, \biggl[
	-
	\frac{2 \overline{\mathcal{K}}}{3} \frac{ \delta \pi }{ \sqrt{ \overline{h} } }
	+
	\frac{1}{18} 
		\Bigl( \overline{\mathcal{K}}^2 
			\!-\! 3\overline{\mathcal{K}}{}^{ij} \overline{\mathcal{K}}_{ij} 
			\!+\! 3\overline{\mathcal{R}} \Bigr)
		\frac{ \delta\rho }{ \sqrt{ \overline{h} } }
	-
	\frac{1}{3}
	\overline{\nabla}{}^i \overline{\nabla}_i 
		\Bigl( \frac{ \delta\rho }{ \sqrt{ \overline{h} } } \Bigr)
	\biggr]
	\, ,
\\
\mathcal{H}^i_{\scr (1)} \approx{}&
	\sqrt{\overline{h}} \, \biggl[
	-
	\frac{2}{3} \overline{\nabla}{}^i \Bigl( \frac{ \delta \pi }{ \sqrt{ \overline{h} } } \Bigr)
	-
	\frac{\overline{\mathcal{K}}}{9} \overline{\nabla}{}^i \Bigl( \frac{ \delta\rho }{ \sqrt{ \overline{h} } } \Bigr)
	+
	\frac{2}{9} \frac{ \delta\rho}{ \sqrt{ \overline{h} } } 
		\overline{\nabla}{}^i \overline{\mathcal{K}}
	-
	\frac{1}{3} \overline{\nabla}_j \Bigl( \overline{\mathcal{K}}{}^{ij} 
		\frac{ \delta\rho }{ \sqrt{ \overline{h} }} \Bigr)
	\biggr]
	\, .
\end{align}
This way the formally four first-class constraints depend on the two scalar perturbations only.
Because of this they have to be counted as only two first-class constraints, and essentially
reduce to~$\delta\rho\approx 0$ and~$\delta \pi \approx 0$. This brings the count of the 
total number of first-class constraints to~$N_{\rm 1st}=12$, and consequently the number of 
physical propagating degrees of freedom to zero,
\begin{equation}
N_{\rm phy} = \frac{1}{2} \Bigl( 24 - 2\times 12 - 0 \Bigr) = 0 \, .
\end{equation}
%

%%%%%%%%%%%%%%%%%%%%%%%%%%%%%%%%%%%%%%%%%%%%%%%
%%%	ACCIDENTAL GAUGE SYMMETRY
%%%%%%%%%%%%%%%%%%%%%%%%%%%%%%%%%%%%%%%%%%%%%%%
\paragraph{Accidental gauge symmetry.}
The change of character of traceless constraints from second-class to first-class, and the loss 
of transverse components of the momentum constraint suggest that the structure of local 
symmetries of the theory have been modified by the linearisation process.
Indeed, the singularity of traceless-Ricci backgrounds in the Lagrangian 
formulation for linearised perturbations,
\begin{equation}
S[\delta g_{\mu\nu}] = \int\! d^4x \, \sqrt{-\overline{g}} \,
	\Bigl[ \Bigl( \overline{D}{}^\mu \overline{D}{}^\nu
		- \overline{g}{}^{\mu\nu} \overline{D}{}^\rho \overline{D}_\rho
		- \overline{R}{}^{\mu\nu} \Bigr) \delta g_{\mu\nu} \Bigr]^2
	\, ,
\end{equation}
manifests itself through an accidental gauge symmetry for linear perturbations, 
\begin{equation}
\delta g_{\mu\nu} \longrightarrow \delta g_{\mu\nu}
	+
	\mathcal{P}_{\mu\nu}{}^{\rho\sigma} \xi_{\rho\sigma} \, ,
\label{GaugeTrans}
\end{equation}
where~$\xi_{\mu\nu}$ is an arbitrary symmetric 2-tensor field, 
and~$\mathcal{P}_{\mu\nu}{}^{\rho\sigma}$ is the appropriate projector.

While finding the projector in~(\ref{GaugeTrans}) explicitly is generally not a 
straightforward task,
it is not difficult to construct it for more special Ricci-flat backgrounds, 
where~$\overline{R}_{\mu\nu}\!=\!0$. 
This is accomplished in Appendix~\ref{sec: Perturbing Bach tensor} with the help of 
the Bach tensor, that is both transverse and traceless on arbitrary backgrounds, and 
vanishes for Ricci-flat backgrounds. The projector can be read off from the linear 
perturbation of the Bach tensor around Ricci-flat backgrounds,
and written in a more compact form upon commuting some covariant derivatives,
\begin{align}
\mathcal{P}_{\mu\nu}{}^{\rho\sigma}
	={}&
	\Pi^{(\rho}{}_{(\mu} \Pi^{\sigma)}{}_{\nu)}
		- \frac{1}{3} \Pi_{(\mu\nu)} \Pi^{(\rho\sigma)}
	+
	\overline{R}_{(\mu}{}^\alpha{}_{\nu)}{}^{(\rho} \Pi^{\sigma)}{}_\alpha
	+
	\frac{1}{2} \overline{R}_{(\mu}{}^\alpha{}_{\nu)}{}^{\beta}
		\overline{R}_{\alpha}{}^{(\rho}{}_\beta{}^{\sigma)}
	-
	\overline{D}^\alpha \overline{D}_{(\mu} \overline{R}_{\nu)}{}^{(\rho}{}_\alpha{}^{\sigma)}
	\, ,
\label{projector}
\end{align}
where
\begin{equation}
\Pi_{\mu\nu} = \overline{g}_{\mu\nu} \overline{D}{}^\alpha \overline{D}_{\alpha}
	- \overline{D}_{\mu} \overline{D}_{\nu}
\end{equation}
is a transverse projector when acting on a vector, and where derivatives act on everything 
to the right of them. The projector in~(\ref{projector})
is transverse and traceless when contracted onto a 2-tensor, so that the transformation
in~(\ref{GaugeTrans}) is guaranteed to be a symmetry for an arbitrary symmetric
tensor~$\xi_{\mu\nu}$. Furthermore, it reduces to the transverse-traceless projector in 
flat space, and reproduces the accidental gauge symmetry found 
in~\cite{Karananas:2024hoh}.

\paragraph{Nonlinear perturbations.}
While the spectrum of linear perturbations around traceless-Ricci backgrounds is empty,
it is not immediately clear what the conclusion would be at higher order in perturbation theory.
This question can be addressed in the same manner as for the special case of Minkowski 
background in Sec.~\ref{subsec: Nonlinear perturbations}, by assuming a power series
expansion for all variables around the background,
\begin{subequations}
\begin{align}
h_{ij} \approx{}&
	\overline{h}_{ij} 
	+ \varepsilon \delta \overline{h}{}^1_{ij} 
	+ \varepsilon^2 \delta \overline{h}{}^2_{ij} 
	+ \varepsilon^3 \delta \overline{h}{}^3_{ij} 
	+ \dots
	\, ,
\\
\pi^{ij} \approx{}& 
	0
	+ \varepsilon \delta \overline{\pi}{}_1^{ij} 
	+ \varepsilon^2 \delta \overline{\pi}{}_2^{ij} 
	+ \varepsilon^3 \delta \overline{\pi}{}_3^{ij} 
	+ \dots
	\, ,
\\
\mathcal{K}_{ij} \approx{}&
	\overline{\mathcal{K}}_{ij}
	+ \varepsilon \delta \overline{\mathcal{K}}{}^1_{ij} 
	+ \varepsilon^2 \delta \overline{\mathcal{K}}{}^2_{ij} 
	+ \varepsilon^3 \delta \overline{\mathcal{K}}{}^3_{ij} 
	+ \dots
	\, ,
\\
\rho^{ij} \approx{}& 
	0
	+ \varepsilon \delta \overline{\rho}{}_1^{ij} 
	+ \varepsilon^2 \delta \overline{\rho}{}_2^{ij} 
	+ \varepsilon^3 \delta \overline{\rho}{}_3^{ij} 
	+ \dots
	\, ,
\\
N \approx{}& 
	\overline{N}
	+ \varepsilon  \delta \overline{N}_1
	+ \varepsilon^2  \delta \overline{N}_2 
	+ \varepsilon^3  \delta \overline{N}_3
	+ \dots
	\, ,
\\
N_i \approx{}&
	\overline{N}_i
	+ \varepsilon  \delta \overline{N}{}_1^i
	+ \varepsilon^3  \delta \overline{N}{}_2^i
	+ \varepsilon^3  \delta \overline{N}{}_3^i
	+ \dots
	\, ,
\\
\lambda_{ij} \approx{}& 
	\overline{\lambda}_{ij}
	+ \varepsilon \delta \overline{\lambda}{}^1_{ij} 
	+ \varepsilon^2 \delta \overline{\lambda}{}^2_{ij} 
	+ \varepsilon^3 \delta \overline{\lambda}{}^3_{ij} 
	+ \dots
	\, .
\end{align}
\end{subequations}
where the bookkeeping parameter~$\varepsilon$ keeps track of the powers of
perturbation fields.

We will see that perturbation theory fails to capture any propagating
degrees of freedom around all traceless-Ricci backgrounds. Proving this is analogous,
though much more tedious, to the procedure in Sec.~\ref{subsec: Nonlinear perturbations}.
It relies on the result from Appendix~\ref{sec: Higher order perturbations} which tells 
us that the action at each successive perturbative order remains quadratic,
\begin{align}
\MoveEqLeft[2.5]
\mathscr{S}_n \bigl[ \delta N, \delta N_i, \delta \lambda_{ij}, \delta h_{ij}, 
	\delta \pi^{ij}, \delta \mathcal{K}_{ij}, \delta \rho^{ij} \bigr] 
\nonumber \\
	={}&
	\varepsilon^{2n} \!\! \int\! d^4x \, \Bigl[
		\delta \pi^{ij} \delta \dot{h}_{ij}
		+
		\delta \rho^{ij} \delta \dot{\mathcal{K}}_{ij}
		-
		\mathscr{H}_n
		-
		\delta N \mathcal{H}_{\scr (1)}
		-
		\delta N_i \mathcal{H}^i_{\scr (1)}
		-
		\delta \lambda_{ij} \Phi^{ij}_{\scr (1)}
		\Bigr]
		+
		\mathcal{O}(\varepsilon^{2n+1})
		\, ,
\end{align}
such that the constraints, given in~(\ref{HamiltonianLinear})--(\ref{PhiLinear}), remain the same,
and that the Hamiltonian density,
\begin{equation}
\mathscr{H}_n = \mathscr{H}_{\scr (2)} + \Delta\mathscr{H}_n \, ,
\end{equation}
contains the quadratic part~$\mathscr{H}_{\scr (2)}$, given in~(\ref{secondH}), that does not
change, and a linear part~$\Delta\mathscr{H}_n$ that changes from order to order.
Furthermore, the perturbations of the momenta~$\delta \pi^{ij}$ and~$\delta \rho^{ij}$
vanish on-shell at every order, so the linear part of the Hamiltonian has to take the form
\begin{equation}
\Delta\mathscr{H}_n = A_n \delta \rho + B_{ij}^n \delta \pi^{ij} \, ,
\end{equation}
where it is only the coefficient functions~$A_n$ and~$B^n_{ij}$ that are updated at 
each order. This is why the linear part of the Hamiltonian density does not influence 
the constraint structure, that remains the same at each order. Consequently, we find 
no propagating degrees of freedom at each perturbative order.

Attention should be paid to the detail that primary constraints are now explicitly 
time-dependent, but only through the background field dependence. This time dependence, 
that is the same at each order, has to be accounted for when demanding conservation of 
constraints.

Therefore, the degenerate constraint structure for perturbations around traceless-Ricci
backgrounds does not change at any order, and consistently yields zero propagating
degrees of freedom. However, absence of degrees of freedom is not a physical feature of 
the vicinity of traceless-Ricci spacetimes. The interpretation of this feature is the same
as in the Minkowski case from Sec.~(\ref{subsec: Nonlinear perturbations}): perturbation 
theory is not appropriate for examining the vicinity of traceless-Ricci spacetimes. The dynamics of the 
variable~$\delta \rho$, that measures departure from the vanishing Ricci scalar,
becomes nonperturbative close to traceless-Ricci spacetimes, and different methods
have to be used to quantify this region in field space.

%%%%%%%%%%%%%%%%%%%%%%%%%%%%%%%%%%%%%%%%%%%%%%%
%%%%%%%%%%%%%%%%%%%%%%%%%%%%%%%%%%%%%%%%%%%%%%%
%%%%%%%%%%%%%%%%%%%%%%%%%%%%%%%%%%%%%%%%%%%%%%%
%%%	COSMOLOGICAL PHASE SPACE
%%%%%%%%%%%%%%%%%%%%%%%%%%%%%%%%%%%%%%%%%%%%%%%
%%%%%%%%%%%%%%%%%%%%%%%%%%%%%%%%%%%%%%%%%%%%%%%
%%%%%%%%%%%%%%%%%%%%%%%%%%%%%%%%%%%%%%%%%%%%%%%
\section{Cosmological phase space}
\label{sec: Cosmological phase space}

Examples of singular spacetimes that exhibit the strong coupling feature given in 
Sec.~\ref{sec: Perturbations around other singular backgrounds} are all spacetimes 
with globally and eternally vanishing Ricci scalar. Slight deviations from such spacetimes 
cannot be described by perturbation theory. This begs the question of whether such 
backgrounds can be reached dynamically, or whether they are in some sense isolated 
from the rest of the phase space. In order to address this question, at least in part, 
in this section we give the dynamical system analysis for spatially flat cosmology of 
pure~$R^2$ theory. We find that the~$R=0$ point is indeed crossed by some trajectories 
of the evolution. We focus on demonstrating this by providing various phase space plots,
and consequently we do not delve into analysing various attractors and 
singular points of the phase space flow.

It might be tempting to perform this analysis in the Einstein frame. However, 
we refrain from this on account of the singularities introduced by the conformal 
transformation itself~\cite{Alho:2016gzi}. Such singularities are located precisely at 
points~$R\!=\!0$ that we seek to explore. For this reason we perform the dynamical 
systems analysis directly in the Jordan frame. Even then, there remain many different 
ways to formulate a dynamical system~\cite{Carloni:2006mr,deSouza:2007zpn,Carloni:2007eu,Carloni:2007br,Odintsov:2017tbc,Chakraborty:2021mcf}. The case at hand is 
most similar to the~$f(R)\!=\! R^n$ theory studied in~\cite{Carloni:2004kp}. However, 
the variables used there were developed for a different purpose, and cannot be used 
in practice to probe the strongly coupled region that we are interested in. Accordingly, 
here we take a different approach.

The line element for spatially flat FLRW spacetime was already given in~(\ref{FLRWline}). 
For this spacetime the Ricci tensor is diagonal,
\begin{equation}
R_{00} = - 3 \bigl( \dot{H} + H^2 \bigr)
	\, ,
\qquad \quad
R_{ij} = a^2 \delta_{ij} \bigl( \dot{H} + 3 H^2 \bigr)
	\, .
\end{equation}
while the  the Ricci scalar reads
\begin{equation}
R = 6 \bigl( \dot{H} + 2 H^2 \bigr) \, .
\label{RicciScalar}
\end{equation}
Equations of motion~(\ref{R2eom}) are also diagonal, and reduce to the two Friedmann 
equations,
\begin{equation}
H \dot{R} + H^2 R - \frac{R^2}{12} = 0 \, ,
\qquad \quad
\ddot{R} + 2 H \dot{R} - H^2 R + \frac{R^2}{12} = 0 \, .
\label{FLRWeinstein}
\end{equation}

The three equations in~(\ref{RicciScalar}) and~(\ref{FLRWeinstein}) form the basis
of the dynamical system formulation. However, not all three equations are 
independent. For instance, the second Friedmann equation in~(\ref{FLRWeinstein}) can 
be derived from the first Friedmann equation, and the definition of the Ricci 
scalar~(\ref{RicciScalar}).  In the remainder of this section we first give three different 
two-dimensional dynamical system formulations, analogous to the ones considered
in~\cite{Capozziello:1993xn}, with the goal of illustrating the dynamics 
from multiple perspectives. To this end, it is convenient to define dimensionless quantities,
\begin{equation}
X = H/\kappa \, ,
\qquad\quad
Y = R / \kappa^2 \, ,
\qquad\quad
Z = \dot{R} / \kappa^3 \, ,
\end{equation}
where~$\kappa$ is some arbitrary scale. It is also convenient to rescale time with the 
same scale,~$T \!=\! \kappa t$, which makes the problem dimensionless. The phase 
space is broken up into four distinct partitions in two-dimensional formulations of the 
dynamical system. These are uniquely characterised by the sign of the Hubble rate, and 
the sign of the velocity of the Ricci scalar, that never change during the evolution. 
Color coding of the partitions, that we use in figures throughout the section, is given 
in Table~\ref{Sectors} below. We conclude the section with a one-dimensional 
formulation that clearly reveals that evolution can cross the point~$R=0$.
%%%%%%%%%%%%%%%%%%%%%
\begin{table}[h!]
\vskip+5mm
\renewcommand{\arraystretch}{1.6}
\setlength{\tabcolsep}{1.5em}
\centering
\begin{tabular}{c c c c c}
\hline
	${\rm sgn}(H)$	&	\Large$\boldsymbol+$	&	\Large$\boldsymbol+$	
	&	\Large$\boldsymbol-$	&	\Large$\boldsymbol-$
\\
\hline
${\rm sgn}(\dot{R})$	&	\Large$\boldsymbol+$	&	\Large$\boldsymbol-$	
	&	\Large$\boldsymbol+$	&	\Large$\boldsymbol-$
\\
\hline
colour	&	$\vcenter{\hbox{\includegraphics[width=1cm]{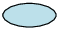}}}$	
	&	$\vcenter{\hbox{\includegraphics{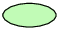}}}$
	&	$\vcenter{\hbox{\includegraphics{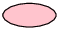}}}$
	&	$\vcenter{\hbox{\includegraphics{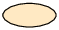}}}$
\\
\hline
\end{tabular}
\caption{Four partitions of the cosmological evolution in the pure~$R^2$ theory,
classified by the sign of the Hubble rate, and the sign of the velocity of the Ricci scalar.}
\label{Sectors}
\end{table}
%%%%%%%%%%%%%%%%%%%%%

\begin{figure}[h!]
\center
\vskip-10mm
\includegraphics[width=9.3cm]{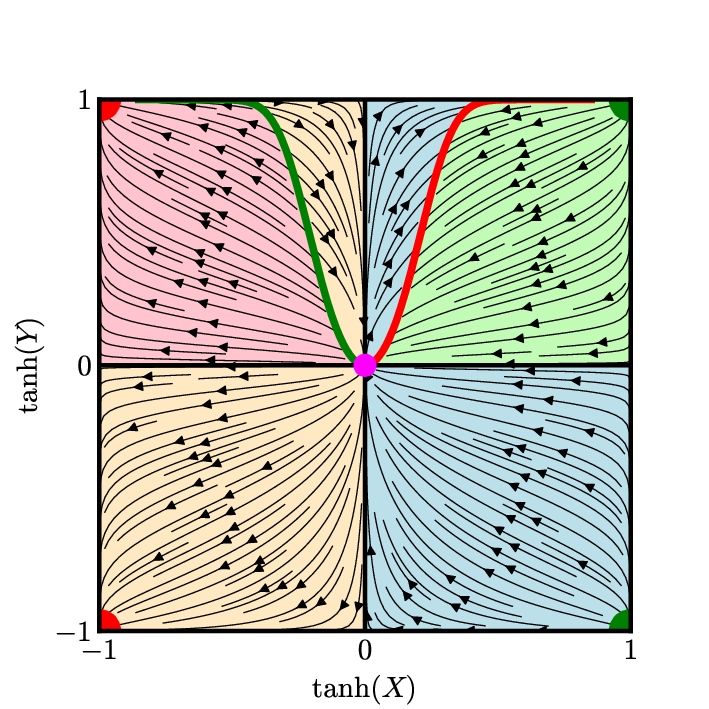}\hfill
\vskip-2mm
\caption{
Compactified phase space flow for the system of autonomous equations 
in~(\ref{DynSys1}). The four sectors of evolution are indicated in different pastel colors 
defined in Table~\ref{Sectors}. The stream lines in green and blue partitions originate 
in respective sources indicated by green dots in the corners of the plot, and terminate at the
red curve representing an expanding de Sitter attractor; the stream lines in red and yellow 
partitions originate at the green curve representing a contracting de Sitter repulsor, and 
terminate at respective sinks indicated by red dots in corners.
Sources and sinks appear confined at the corners, as a result of our hyperbolic 
compactification. 
The lines at~$X\!=\!0$ and~$Y\!=\!0$ partition the flow at all points except for 
the magenta point at the origin--- at this point we claim that the blue and green evolutions penetrate the 
strongly coupled surface. The upper-right quadrant matches that given 
in~\cite{Chakraborty:2021mcf}.}
\label{DSplot1}
\end{figure}

%%%%%%%%%%%%%%%%%%%%%%%%%%%%%%%%%%%%%%%%%%%%%%%
%%%	FIRST TWO--DIMENSIONAL FORMULATION
%%%%%%%%%%%%%%%%%%%%%%%%%%%%%%%%%%%%%%%%%%%%%%%
\paragraph{First two-dimensional formulation.}
In the first representation of the system, we simply take the definition of the Ricci scalar and the first Friedmann equation as independent equations, which in dimensionless variables read
\begin{equation}
\frac{dX}{dT} = \frac{Y}{6} - 2 X^2 \, ,
\qquad \quad
\frac{dY}{dT} = \frac{Y^2}{12X} - XY \, .
\label{DynSys1}
\end{equation}
The compactified phase space plot following from these equations is given in Fig.~\ref{DSplot1}. 
The autonomous system in~(\ref{DynSys1}) has the advantage of being described by rational 
functions. Accordingly, almost the whole volume of the infinite two-dimensional plane 
corresponds to a well-defined flow, with the exceptions arising along~$X=0$, which is a set of 
measure zero. Four distinct cosmological evolutions are visible, and in line with expectation 
that the strongly coupled surface~$Y\!=\!0$ mostly manifests as a separatrix to partition these. 
This partitioning appears to break down, however, at precisely the 
point~$X\!=\!Y\!=\!0$ where the system is no longer faithful to the physics. As illustrated 
in Fig.~\ref{DSplot1}, we actually claim that two of the four evolutions penetrate through the 
strongly coupled surface at this point: since the point in question involves the convergence of 
infinitely many flow lines, and a corresponding loss of information, we must transition to 
improved coordinates to defend this claim. It will be clear that this feature is a 
consequence of the two-dimensional projection of what is in essence a two-dimensional 
curved surface embedded in a three-dimensional phase space.
%%%%%%%%%%%%%%%%%%%%%%
%%%%%%%%%%%%%%%%%%%%%%

\begin{figure}[h!]
\center
\includegraphics[width=16cm]{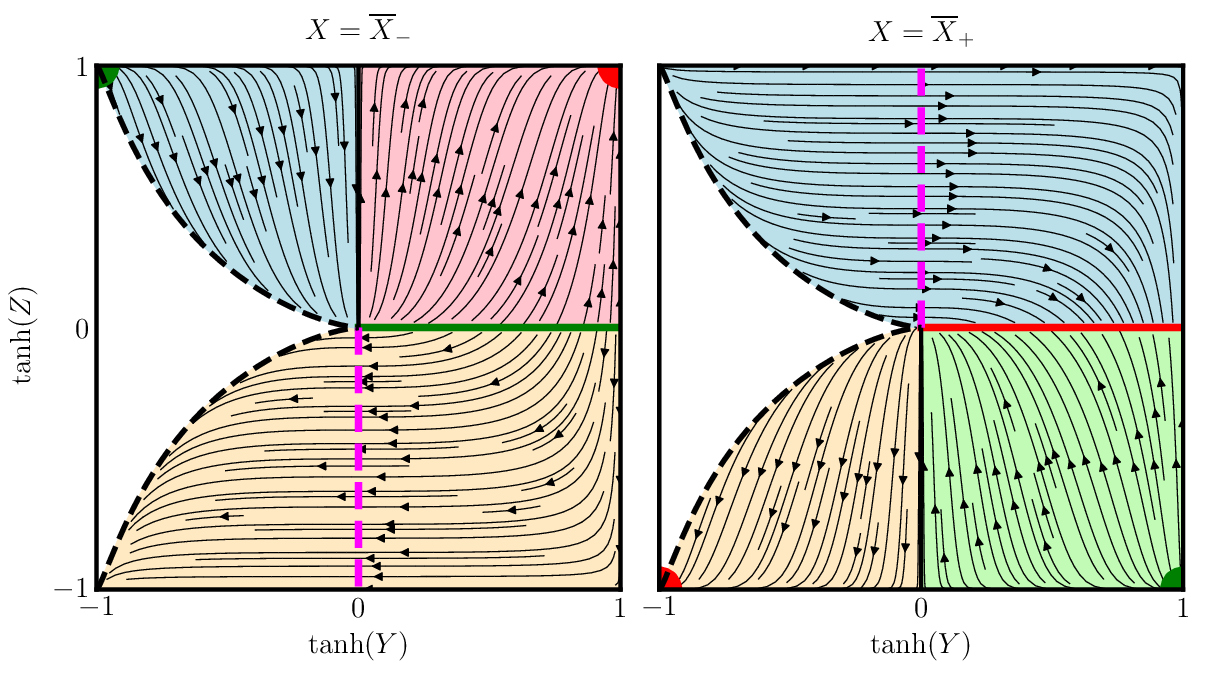}
\vskip-2mm
\caption{
Compactified phase space flow for the system of autonomous equations 
in~(\ref{DynSys2}), producing an alternative perspective to that shown in 
Fig.~\ref{DSplot1}. In these coordinates, the blue and yellow evolutions are 
split over the two branches, 
which must be glued along the matching cuts indicated by black dashed lines. The flow
lines bounce tangentially from these dashed lines when transitioning from one
branch to another. Both blue and yellow evolutions clearly pass through the strongly 
coupled surface at~$Y\!=\!0$, indicated by the dashed magenta lines, which degenerate 
into the single magenta point at the origin in Fig.~\ref{DSplot1}. The de Sitter repulsor 
(green) and attractor (red) lie along the line~$Z\!=\!0$ for~$Y\!>\!0$. The sources and 
sinks in the corners are indicated by green and red dots, respectively.}
\label{DSplot2}
\end{figure}

%%%%%%%%%%%%%%%%%%%%%%%%%%%%%%%%%%%%%%%%%%%%%%%
%%%	SECOND TWO--DIMENSIONAL FORMULATION
%%%%%%%%%%%%%%%%%%%%%%%%%%%%%%%%%%%%%%%%%%%%%%%
\paragraph{Second two-dimensional formulation.}
To more thoroughly explore what happens at the origin of Fig.~\ref{DSplot1}, we
choose the Ricci scalar and its velocity to span the phase space. In order to do this we 
need to solve for the Hubble rate from the first Friedmann equation in terms of the 
other two variables, so as to eliminate it algebraically from the system. In terms of the 
variable~$X$, the first Friedmann equation is a quadratic that produces two branches
\begin{equation}
X = \overline{X}_{\pm}(Y,Z) \equiv \frac{1}{2Y} \biggl[ - Z \pm \sqrt{ \frac{3Z^2+Y^3}{3} } \ \biggr] \, .
\end{equation}
Note that the discriminant excludes a part of the phase space for 
which~$3Z^2 \!+\! Y^3 \!<\! 0$. We plug this solution into the remaining equations 
to form the dynamical system,
\begin{equation}
\frac{dY}{dT} = Z \, ,
\qquad\quad
\frac{dZ}{dT} = - 3 \overline{X}_{\pm} Z \, .
\label{DynSys2}
\end{equation}
Note that the first equation is the trivial definition of the Ricci scalar derivative, so that
only the second equation is sensitive to the branching.
The corresponding plot of the phase space flow is given in Fig.~\ref{DSplot2}. Despite the 
mild complication of having to glue the branches along black curved dashed cuts where the 
discriminant vanishes, the essential point is clearly visible, that the blue and yellow 
evolutions penetrate the strongly coupled surface at~$Y\!=\!0$, indicated by the 
dashed magenta line.
%%%%%%%%%%%%%%%%%%%%%%
%%%%%%%%%%%%%%%%%%%%%%

\begin{figure}[h!]
\center
\includegraphics[width=16cm]{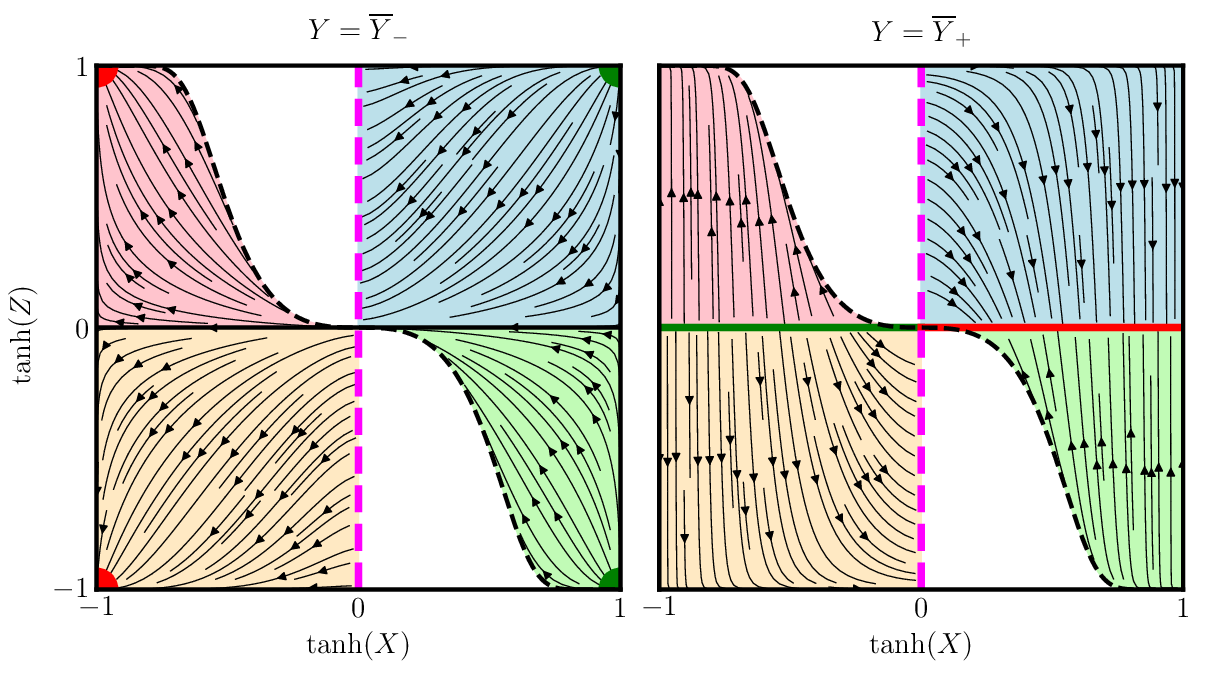}
\vskip-2mm
\caption{
Compactified phase space flow for the system of autonomous equations in~(\ref{DynSys3}),
producing an alternative perspective to that shown in Figs.~\ref{DSplot1} and~\ref{DSplot2}.
In these coordinates, all evolutions are split over the two branches, which must be glued 
along the matching cuts indicated by dashed lines. The flow lines bounce 
tangentially from the black dashed lines, and continue orthogonally through the magenta
dashed lines, that represent the singular surface of the vanishing Ricci scalar. Sources and 
sinks in corners on the left plot are indicated by green and red dots, respectively, while
de Sitter repulsor and attractor on the right plot are indicated by the green and red lines,
respectively.
}
\label{DSplot3}
\end{figure}

%%%%%%%%%%%%%%%%%%%%%%%%%%%%%%%%%%%%%%%%%%%%%%%
%%%	THIRDTWO--DIMENSIONAL FORMULATION
%%%%%%%%%%%%%%%%%%%%%%%%%%%%%%%%%%%%%%%%%%%%%%%
\paragraph{Third two-dimensional formulation.}
The last combination of quantities we can use to form a two-dimensional phase space
flow are the Hubble rate and the Ricci scalar velocity. In this case we again solve the 
first Friedmann equation algebraically, but this time for the Ricci scalar,
\begin{equation}
Y = \overline{Y}_{\pm}(X,Z) \equiv 
	6\biggl[
	X^2 \pm \sqrt{ \frac{ X (3 X^3 \!+\! Z ) }{3} } \
	\biggr]
	\, .
\end{equation}
The discriminant again cuts out part of the phase space for which~$X (3 X^3 \!+\! Z ) \!<\! 0$.
Since this discriminant factorises, the result will be an awkward pair of cuts.  Plugging the
solution into the definition of the Ricci scalar and the second Friedmann equation then
gives the two equations of the dynamical system,
\begin{equation}
\frac{dX}{dT} = 	\frac{\overline{Y}_\pm}{6} - 2 X^2 \, ,
\qquad \quad
\frac{dZ}{dT} = - 3 X Z
\, .
\label{DynSys3}
\end{equation}
The plot of the phase space flow is given in figure~\ref{DSplot3}. This is provided largely out of 
completeness, since the essential feature of penetration through the strongly coupled 
surface is already clear from Figs.~\ref{DSplot1} and~\ref{DSplot2}.
%%%%%%%%%%%%%%%%%%%%%%
%%%%%%%%%%%%%%%%%%%%%%

\begin{figure}[h!]
\center
\includegraphics[width=9.3cm]{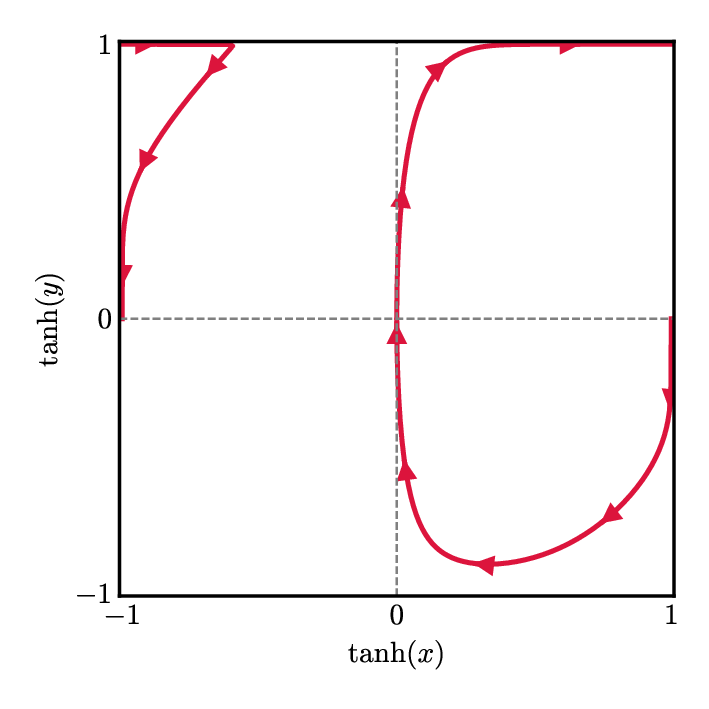}
\vskip-5mm
\caption{
Compactified flow trajectory for the system in~(\ref{DynSys1D}). The direction 
of the flow is with respect to the dimensionless time defined in~(\ref{TauTime}).
However, note that the dimensionful time follows this flow if the Ricci scalar
velocity is positive,~$\dot{R}\!>\!0$, and actually flows in reverse direction
when~$\dot{R}\!<\!0$. This is the reason why the right curve captures both
the blue and the yellow sectors defined in Table~\ref{Sectors}, and the left curve
captures the green and red sectors. For sectors with positive Ricci velocity
the evolution in dimensionful time follows the clockwise direction indicated, while for 
sectors with negative Ricci velocity this flow is anti-clockwise.
The penetration through the strongly coupled surface
at~$y\!=\!0$ is evident for the right curve, that captures both blue and yellow partitions
from Figs.~\ref{DSplot1}--\ref{DSplot3}. }
\label{PhaseSpace1D}
\end{figure}

%%%%%%%%%%%%%%%%%%%%%%%%%%%%%%%%%%%%%%%%%%%%%%%
%%%	ONE-DIMENSIONAL FORMULATION
%%%%%%%%%%%%%%%%%%%%%%%%%%%%%%%%%%%%%%%%%%%%%%%
\paragraph{One-dimensional formulation.}
Finally, we present an alternative to the three 
two-dimensional dynamical system formulations 
presented above. Given that none of the four partitions cross the~$\dot{R}\!=\!0$
point, we may adopt~$\dot{R}$ as the natural scale of the problem.\footnote{
One should be careful not to introduce singular points in the dynamical system
that are just the property of the variables used, as opposed to the genuine singular 
points of the evolution~\cite{Odintsov:2018uaw}.
}
Therefore, we can adopt the Hubble rate and the Ricci scalar as variables in these 
natural units,
\begin{equation}
x = H / \dot{R}^{1/3} = X/Z^{1/3} \, ,
\qquad \quad
y = R / \dot{R}^{2/3}
	= Y / Z^{2/3}  \, .
\end{equation}
Together with the definition of dimensionless time,
\begin{equation}
\tau = t \dot{R}^{1/3} \, ,
\label{TauTime}
\end{equation}
this reduces the three equations~(\ref{RicciScalar}) and~(\ref{FLRWeinstein})
into a single first order equation, supplemented by an algebraic equation,
\begin{equation}
\frac{dx}{d\tau} = - x^2 + \frac{y}{6}
	\, ,
\qquad \quad
x + x^2 y - \frac{y^2}{12} = 0 \, ,
\label{DynSys1D}
\end{equation}
with the third equation becoming redundant.
The effectively one-dimensional dynamical system formulation in~(\ref{DynSys1D}) is 
possible for monomial~$f(R)\!=\!R^n$, but not for general~$f(R)$ theories.
Fig.~\ref{PhaseSpace1D} gives the solutions of this system,
represented by the two disjoint curves. The direction of flow along these two curves is 
either clockwise or anti-clockwise, and is determined only when the sign of the Ricci scalar 
velocity is chosen. In this way the curve that passes through the~$R\!=\!0$ point corresponds 
to blue and yellow partitions from two-dimensional formulations, and the curve that does
not pass through that point corresponds to red and green partitions. Once again, we
clearly see that the points at which~$R\!=\!0$ can be traversed during the cosmological
evolution.
%

%%%%%%%%%%%%%%%%%%%%%%%%%%%%%%%%%%%%%%%%%%%%%%%
%%%%%%%%%%%%%%%%%%%%%%%%%%%%%%%%%%%%%%%%%%%%%%%
%%%%%%%%%%%%%%%%%%%%%%%%%%%%%%%%%%%%%%%%%%%%%%%
%%%	DISCUSSION
%%%%%%%%%%%%%%%%%%%%%%%%%%%%%%%%%%%%%%%%%%%%%%%
%%%%%%%%%%%%%%%%%%%%%%%%%%%%%%%%%%%%%%%%%%%%%%%
%%%%%%%%%%%%%%%%%%%%%%%%%%%%%%%%%%%%%%%%%%%%%%%
\section{Discussion}
\label{sec: Discussion}

In this work, we have performed a full Hamiltonian constraint analysis of 
pure~$R^2$ theory of gravity in order to address recent controversies regarding its 
particle spectrum. Our analysis confirms that in general the full theory propagates three degrees of 
freedom, corresponding to a massive spin-two graviton and a scalar, in agreement with 
established results~\cite{Buchbinder:1987vp}. However, the constraint structure 
reveals singular points in field space where constraints change character. This feature is
behind the reported absence of propagating linearised perturbations around Minkowski 
space~\cite{Hell:2023mph,Golovnev:2023zen,Karananas:2024hoh}~\footnote{
The absence of linearised degrees of freedom around Minkowski space has also
been shown in~$D$ spacetime dimensions~\cite{Hell:2025wha}.}. 

Furthermore, we have 
shown that the absence of propagating modes in the linearised theory 
is not a phenomenon specific to Minkowski space, but rather a generic feature of any 
background with a vanishing Ricci scalar, $R\!= \!0$, i.e.~traceless-Ricci background. 
The perturbative analysis fails on these
singular surfaces because the constraint algebra is fundamentally altered: ten second class
constraints of the full theory become first class, while the momentum constraint becomes
degenerate as it loses its transverse part. This feature is responsible for removing all 
degrees of freedom from the linearised spectrum. Moreover, considering higher order
perturbation theory around such surfaces, organized in powers of the perturbations,
yields no propagating degrees of freedom at every order. While such a perturbation theory 
ought to probe the vicinity of traceless-Ricci spacetimes, it is in conflict with the
general result of three propagating degrees of freedom that is valid arbitrarily
close to traceless-Ricci spaces.
Thus, this result should be interpreted as a limitation of the perturbation theory,
and its inability to describe this regime of the theory.  Rather, the dynamics of
perturbations should become nonperturbative in this regime.

It is worth pointing out that a similar issue has also been reported for the Palatini
(metric-affine) formulation of pure~$R^2$ theory, where no degrees of freedom are found 
in the linearised spectrum around Minkowski spacetime~\cite{Karananas:2024qrz}. 
In that context, the full theory is known to propagate just the two degrees of freedom of 
the graviton, a result which has also been confirmed by Hamiltonian 
analysis~\cite{Glavan:2023cuy}. The recurrence of this pathology underscores
that the surfaces where the theory becomes strongly coupled are a key, non-trivial feature
of higher-derivative gravity models. Similar features where a subset of degrees of
freedom disappears from the linearised spectrum have been observed for the
cuscuton model~\cite{Gomes:2017tzd}, and also for the Einsteinian Cubic Gravity and its 
generalizations~\cite{BeltranJimenez:2020lee}. In the latter instance it was found that
these singular surfaces in field space are shielded from the rest of the phase space by the
nonperturbative behaviour of degrees of freedom in their vicinity. This begs the question 
whether something like this happens in the pure~$R^2$ model we consider here.

The cosmological phase-space analysis we give in Sec.~\ref{sec: Cosmological phase space}
demonstrates that the singular $R\!=\!0$ surface is dynamically accessible. This implies 
that the strong-coupling feature is not merely a mathematical curiosity confined to static, 
eternal spacetimes, but is of direct physical relevance for evolving cosmologies. This opens 
several avenues for future investigation. First, it is unclear what happens to perturbations as 
the cosmological background evolves smoothly through an $R\!=\!0$ phase. We believe this 
question requires a dedicated study using full numerical evolution to properly capture the 
non-linear dynamics.

Second, we conjecture that this phenomenon is a general feature of all~$f (R)$ theories,
manifesting on backgrounds where $f'(R)\!=\!0$. There is already some indication of this in
the literature~\cite{Casado-Turrion:2024esi}, and a full Hamiltonian analysis would be the 
appropriate tool to verify it. Confirming this conjecture would provide a unified 
understanding of the connection between theories like~$R\!+\!R^2$ 
gravity and pure~$R^2$ gravity. Rather than being disconnected theories, as has been 
suggested in~\cite{Hell:2023mph} and~\cite{Karananas:2024hoh}, 
their features would be smoothly related, with the primary difference being the location 
of the singular, strong-coupling surface in the phase space.

%%%%%%%%%%%%%%%%%%%%%%%%%%%%%%%%%%%%%%%%%%%%%%%
%%%%%%%%%%%%%%%%%%%%%%%%%%%%%%%%%%%%%%%%%%%%%%%
%%%%%%%%%%%%%%%%%%%%%%%%%%%%%%%%%%%%%%%%%%%%%%%
%%%	ACKNOWLEDGEMENTS
%%%%%%%%%%%%%%%%%%%%%%%%%%%%%%%%%%%%%%%%%%%%%%%
%%%%%%%%%%%%%%%%%%%%%%%%%%%%%%%%%%%%%%%%%%%%%%%
%%%%%%%%%%%%%%%%%%%%%%%%%%%%%%%%%%%%%%%%%%%%%%%
\section*{Acknowledgements}
\label{sec: Acknowledgements}

We are grateful to Sante Carloni for the discussion on dynamical system analysis of~$f(R)$
theories. We are also indebted to Tom Zlosnik for correcting our application of Hamiltonian methods. We are grateful for discussions with Anamaria Hell and Giorgos Karananas.

W.~B.~is grateful for the support of Girton College, Cambridge, Marie Sk{\l}odowska-Curie
Actions and the Institute of Physics of the Czech Academy of Sciences.
D.~G.~was supported by project 24-13079S of the Czech Science Foundation (GA\v{C}R).

Co-funded by the European Union (Physics for Future – Grant Agreement No.~101081515).
Views and opinions expressed are however those of the author(s) only and do not necessarily
reflect those of the European Union or European Research Executive Agency. Neither the
European Union nor the granting authority can be held responsible for them.

%%%%%%%%%%%%%%%%%%%%%%%%%%%%%%%%%%%%%%%%%%
%%%%%%%%%%%%%%%%%%%%%%%%%%%%%%%%%%%%%%%%%%
%%%	APPENDICES
%%%%%%%%%%%%%%%%%%%%%%%%%%%%%%%%%%%%%%%%%%
%%%%%%%%%%%%%%%%%%%%%%%%%%%%%%%%%%%%%%%%%%
\appendix

%%%%%%%%%%%%%%%%%%%%%%%%%%%%%%%%%%%%%%%%%%%%%%%
%%%%%%%%%%%%%%%%%%%%%%%%%%%%%%%%%%%%%%%%%%%%%%%
%%%%%%%%%%%%%%%%%%%%%%%%%%%%%%%%%%%%%%%%%%%%%%%
%%%	PERTURBING BACH TENSOR
%%%%%%%%%%%%%%%%%%%%%%%%%%%%%%%%%%%%%%%%%%%%%%%
%%%%%%%%%%%%%%%%%%%%%%%%%%%%%%%%%%%%%%%%%%%%%%%
%%%%%%%%%%%%%%%%%%%%%%%%%%%%%%%%%%%%%%%%%%%%%%%
\section{Perturbing Bach tensor}
\label{sec: Perturbing Bach tensor}

The Bach tensor for 4-dimensional spacetimes,
\begin{align}
B_{\mu\nu} = W_{\mu\rho\nu\sigma} S^{\rho\sigma}
	+ D^\rho D_\rho S_{\mu\nu}
	- D^\rho D_{(\mu} S_{\nu)\rho}
	\, ,
\end{align}
is defined in terms of the Weyl tensor,
\begin{equation}
W_{\mu\nu\rho\sigma} = 
	R_{\mu\nu\rho\sigma}
	-
	2 g_{\mu] [\rho} R_{\sigma] [\nu}
	+
	\frac{1}{3} R g_{\mu[\rho} g_{\sigma]\nu}
	\, ,
\end{equation}
and the Schouten tensor,
\begin{equation}
S_{\mu\nu} = \frac{1}{2} R_{\mu\nu} - \frac{1}{12} R g_{\mu\nu} \, .
\label{Schouten}
\end{equation}
It is both transverse,~$D^\mu B_{\mu\nu} \!=\! 0$, and 
traceless,~$g^{\mu\nu}B_{\mu\nu} \!=\! 0$,
for arbitrary spacetimes. For Ricci-flat spacetimes it vanishes,~$B_{\mu\nu}=0$, given that 
the Schouten tensor~(\ref{Schouten}) vanishes there. Therefore, we have that the linearised 
perturbation of the Bach tensor around Ricci-flat background,~$R_{ij}\!=\!0$, 
is also traceless and transverse,
\begin{equation}
\delta \bigl( D^\mu B_{\mu\nu} \bigr) = \overline{D}{}^\mu \delta B_{\mu\nu} = 0 \, ,
\qquad\quad
\delta\bigl( g^{\mu\nu} B_{\mu\nu} \bigr) = \overline{g}{}^{\mu\nu} \delta B_{\mu\nu} = 0 \, ,
\end{equation}
where~$\overline{g}_{\mu\nu}$ is the background metric, and~$\overline{D}_\mu$
the covariant derivative with respect to the background metric.
This perturbation of the Bach tensor receives contributions from the perturbation
of the Schouten tensor only,
\begin{equation}
\delta S_{\mu\nu} 
	=
	- \frac{1}{4} \mathbb{P}_{\mu\nu}{}^{\rho\sigma} \delta g_{\rho\sigma}
	\, ,
\end{equation}
defined in terms of the projector
\begin{equation}
\mathbb{P}_{\mu\nu}{}^{\rho\sigma}
	=
	-
	2 \overline{D}{}^{(\rho} \delta^{\sigma)}_{(\mu} \overline{D}_{\nu)}
	+
	\delta^\rho_{(\mu} \delta^\sigma_{\nu)} \overline{D}{}^\alpha \overline{D}_\alpha
	+
	\overline{g}{}^{\rho\sigma} \overline{D}_{(\mu} \overline{D}_{\nu)}
	+
	\frac{ \overline{g}_{\mu\nu} }{3} \Bigl( \overline{D}{}^{(\rho} \overline{D}{}^{\sigma)}
		- \overline{g}{}^{\rho\sigma} \overline{D}{}^\alpha \overline{D}_\alpha \Bigr)
	\, .
\end{equation}
The perturbation of the Bach tensor is then given by
\begin{equation}
\delta B_{\mu\nu} 
	=
	- \frac{1}{4}
	\mathcal{P}_{\mu\nu}{}^{\rho\sigma} \delta g_{\rho\sigma} \, ,
\end{equation}
where the projector is defined as
\begin{equation}
\mathcal{P}_{\mu\nu}{}^{\rho\sigma} 
	=
	\overline{D}{}^\alpha \overline{D}_\alpha \mathbb{P}_{\mu\nu}{}^{\rho\sigma}
	-
	\overline{D}{}^\alpha \overline{D}_{(\mu}
		\mathbb{P}_{\nu)\alpha}{}^{\rho\sigma}
	+
	\overline{R}_{\mu}{}^\alpha{}_\nu{}^\beta \mathbb{P}_{\alpha\beta}{}^{\rho\sigma}
	\, .
\label{AppProject}
\end{equation}
The very convenient property of this projector is that any 2-tensor contracted into it will 
automatically be traceless and transverse on Ricci-flat backgrounds. It is given in a more
compact form in~(\ref{projector}) in the main text.

%%%%%%%%%%%%%%%%%%%%%%%%%%%%%%%%%%%%%%%%%%%%%%%
%%%%%%%%%%%%%%%%%%%%%%%%%%%%%%%%%%%%%%%%%%%%%%%
%%%%%%%%%%%%%%%%%%%%%%%%%%%%%%%%%%%%%%%%%%%%%%%
%%%	HIGHER ORDER PERTURBATIONS
%%%%%%%%%%%%%%%%%%%%%%%%%%%%%%%%%%%%%%%%%%%%%%%
%%%%%%%%%%%%%%%%%%%%%%%%%%%%%%%%%%%%%%%%%%%%%%%
%%%%%%%%%%%%%%%%%%%%%%%%%%%%%%%%%%%%%%%%%%%%%%%
\section{Higher order perturbations}
\label{sec: Higher order perturbations}

When studying small perturbations around certain field configurations the perturbative 
expansion organized in powers of the perturbation fields seems appropriate.
This is often much easier implemented at
the level of equations of motion. However, analyzing theories with constraints is
more conveniently performed at the level of the action. That is why here we outline
perturbation theory organized as an expansion in powers of perturbation fields
adapted for the level of the action. In sections~\ref{subsec: Nonlinear perturbations} 
and~\ref{subsec: General traceless-Ricci spacetimes} we apply this method to 
study higher order perturbations around traceless-Ricci backgrounds in pure~$R^2$ theory.

Consider some action~$S[X]$ that is a functional of the fields~$X^{\tt a}(x)$, where 
index~$\tt a$ labels all the fields and their components. In this appendix we greatly 
benefit from employing DeWitt's shorthand notation, in which the capital Latin index 
labels both the type of the field and its coordinate dependence,~$X^A \!=\! X^{\tt a}(x)$. 
For this appendix it is not relevant whether the action is Lagrangian or Hamiltonian, 
nor what precisely the dynamical fields are. What is important is that the first variation 
of this action generates equations of motion,
\begin{equation}
S_{,A}[\overline{X}] \equiv \frac{\delta S[X]}{\delta X^{\tt a}(x)} \bigg|_{X=\overline{X}} = 0 \, ,
\label{ZerothOrder}
\end{equation}
where we have written the expression using DeWitt's shorthand notation for variational 
derivatives. The last beneficial element of this shorthand notation we need, before 
constructing the perturbative expansion, is the
notation for contracted capital Latin indices,
\begin{equation}
X^A S_{,A}[X] = \int\! d^4x \, \sum_{\tt a} X^{\tt a}(x) \frac{\delta S[X]}{\delta X^{\tt a}(x)} \, ,
\end{equation}
that denotes the sum over field types as well as an integral over coordinate dependence.

Equation~(\ref{ZerothOrder}) defines the background 
solution~$X^A\!\approx\!\overline{X}{}^A$.
We want to derive dynamical equations for small perturbations~$\delta X^A$ around 
that background. Given the assumption that the perturbations are small, the natural 
expansion is in powers of the perturbation fields. That is conveniently kept track 
of by introducing a bookkeeping parameter~$\varepsilon$ that we set to unity at the
end of the computation,~$X^A \!=\! \overline{X}{}^A \!+\!\varepsilon \delta X^A$.
We look for the solution for perturbation fields as a power series in~$\varepsilon$,
\begin{equation}
\varepsilon \delta X^A
	\approx
	\varepsilon \delta \overline{X}{}_1^A
	+
	\varepsilon^2 \delta \overline{X}{}_2^A
	+
	\varepsilon^3 \delta \overline{X}{}_3^A
	+
	\dots
	\, ,
\label{XAexpansion}
\end{equation}
where the coefficient functions~$\delta \overline{X}{}_n^{A}$ are independent 
of~$\varepsilon$. As per usual in perturbation theory, equations of motion for each order  
are derived iteratively, where lower order solutions feed into equations for higher orders.

\paragraph{$\boldsymbol{1}$st order.}
The action for the leading order perturbation is obtained from the starting 
action~$S[X]$ by first shifting the fields by the background 
solution,
\begin{equation}
X^A \longrightarrow \overline{X}{}^A + \varepsilon \delta X^A \, ,
\end{equation}
and plugging this shift into the action.
We then expand the resulting action explicitly in powers of~$\varepsilon$,
keep only the lowest nonvanishing order, and drop the terms independent 
of~$\delta X^A$ that do not contribute to equations of motion,
\begin{equation}
S^1[\delta X]
	\equiv
	S[\overline{X} \!+\! \varepsilon \delta X ] - S[\overline{X}]
	=
	\varepsilon \underbrace{ S_{,A}[\overline{X}] }_{=0} \delta X^A
	+
	\frac{\varepsilon^2}{2} S_{,AB}[\overline{X}]
		\delta X^A \delta X^B
	+ \mathcal{O}(\varepsilon^3)
	\, .
\label{S1}
\end{equation}
The term linear in~$\varepsilon$ vanishes owing to the background equation of
motion~(\ref{ZerothOrder}), which makes the lowest nonvanishing order~$\varepsilon^2$.
Varying this truncated action then generates equations of motion for the first 
perturbative correction~$\delta \overline{X}_1^A$ in~(\ref{XAexpansion}),
\begin{equation}
S^1_{,AB} [\overline{X}] \delta \overline{X}{}_1^B = 0 \, .
\end{equation}
It is convenient to encode the zeroth order and the first order equations
into a single expression,
\begin{equation}
\sum_{k=0}^{1} \frac{\varepsilon^k}{k!}
	\biggl( 
	\frac{\partial^k}{\partial \varepsilon^k}
	S_A[\overline{X} \!+\! \varepsilon \delta \overline{X}_1]
	\,\Big|_{\varepsilon=0} \biggr)
	= 0 \, .
\end{equation}

\paragraph{$\boldsymbol{2}$nd order.}
We proceed in the same manner at second order:
by shifting the fields by the background solution and the first perturbative correction,
\begin{equation}
X^{A} \longrightarrow \overline{X}{}^{A}
	+ \varepsilon \delta \overline{X}{}_1^{A}
	+ \varepsilon^2 \delta X{}^{A}
	\, ,
\end{equation}
and expanding the action accordingly, keeping the lowest nonvanishing order,
\begin{align}
S^2[\delta X]
	\equiv{}&
	S[\overline{X} \!+\! \varepsilon \delta \overline{X}_1 \!+\! \varepsilon^2 \delta X ] 
		- S[\overline{X} \!+\! \varepsilon \delta \overline{X}_1]
	=
	\varepsilon^2 \underbrace{S_{,A}[\overline{X}]}_{=0}
		\delta X^{A}
	+
	\varepsilon^3 \underbrace{S_{, AB}[\overline{X} ]
		\delta \overline{X}{}_1^A}_{=0} \delta X^{ B}
\nonumber \\
&	\hspace{1cm}
	+
	\frac{\varepsilon^4}{2} S_{,ABC}[\overline{X} ]
		\delta \overline{X}{}_1^{A} \delta \overline{X}{}_1^{B} \delta X^{C}
	+
	\frac{\varepsilon^4}{2} S_{,AB}[\overline{X}]
		\delta X^{A} \delta X^{B}
	+ \mathcal{O}(\varepsilon^5)
	\, .
\end{align}
Orders lower than~$\varepsilon^4$ drop out because of lower order equations of motion,
where we again dropped the parts independent of the dynamical variables.
Equation of motion for the second perturbative correction is then obtained 
by varying the truncated action,
\begin{equation}
	S_{,AB}[\overline{X}] \delta \overline{X}{}_2^B
	+
	\frac{1}{2} S_{,ABC}[\overline{X} ]
		\delta \overline{X}{}_1^B \delta \overline{X}{}_1^C
	=
	0
	\, .
\end{equation}
There is again a useful compact form that captures the first three orders of equations of motion,
\begin{equation}
\sum_{k=0}^{2} \frac{\varepsilon^k}{k!}
	\biggl( 
	\frac{\partial^k}{\partial \varepsilon^k}
	S_A[\overline{X} \!+\! \varepsilon \delta \overline{X}_1
		\!+\! \varepsilon^2 \delta \overline{X}_2 ]
	\,\Big|_{\varepsilon=0} \biggr)
	= 0 \, .
\end{equation}

\paragraph{\boldmath $n$-th order.}
The pattern that we see emerging for lower orders continues at each subsequent order.
We can uncover it by following the same steps.
First we shift the variable by solutions for all lower orders,
\begin{equation}
X^A \longrightarrow \overline{X}{}^A
	+ \varepsilon \delta \overline{X}{}_1^A
	+ \dots
	+ \varepsilon^{n-1} \delta \overline{X}{}_{n-1}^A
	 + \varepsilon^n \delta X^A
	 \, ,
\end{equation}
upon which we plug it into the action and expand it to order~$\varepsilon^{2n}$,
\begin{align}
\MoveEqLeft[2]
S^n[\delta X_n]
	=
	S[ \overline{X}
	\!+\! \dots
	\!+\! \varepsilon^{n-1} \delta \overline{X}_{n-1}
	 \!+\! \varepsilon^n \delta X ] 
		- S[ \overline{X}
	\!+\! \dots
	 \!+\! \varepsilon^{n-1} \delta \overline{X}_{n-1} ]
\nonumber \\
={}&
		\frac{\varepsilon^{2n} }{n!}
			\biggl(
			\frac{\partial^n}{\partial \varepsilon^n}
			S_{,A}[\overline{X} 
			\!+\! \dots
			\!+\! \varepsilon^{n-1} \delta \overline{X}_{n-1}]
			\,\Big|_{\varepsilon=0}
			\biggr)
		\delta X{}^A
	+
	\frac{\varepsilon^{2n}}{2} S_{,AB}[\overline{X}]
		\delta X{}^A \delta X{}^B
	+ \mathcal{O}(\varepsilon^{2n+1})
	\, .
\label{nAction}
\end{align}
All the lower orders in this action  vanish because of lower order equations of motion,
that are compactly written as
\begin{equation}
\sum_{k=0}^{n-1} \frac{\varepsilon^k}{k!}
	\biggl( 
	\frac{\partial^k}{\partial \varepsilon^k}
	S_A[\overline{X} \!+\! \varepsilon \delta \overline{X}_1
		\!+\! \dots \!+\! \varepsilon^{n-1} \delta \overline{X}_{n-1} ]
	\,\Big|_{\varepsilon=0} \biggr)
	= 0 \, .
\end{equation}
The equation of motion for the~$n$-th order perturbation is 
obtained by varying the action,
\begin{equation}
	S_{,AB}[\overline{X}] \delta \overline{X}{}_n^B
	+
	\frac{1}{n!}
		\biggl(
		\frac{\partial^n}{\partial \varepsilon^n}
		S_{,A}[\overline{X} 
		\!+\! \dots
		\!+\! \varepsilon^{n-1} \delta \overline{X}_{n-1} ]
		\,\Big|_{\varepsilon=0}
		\biggr)
	=
	0
	\, ,
\end{equation}
that can be written concisely together with all other lower order equations as
\begin{equation}
\sum_{k=0}^{n} \frac{\varepsilon^k}{k!}
	\biggl( 
	\frac{\partial^k}{\partial \varepsilon^k}
	S_A[\overline{X} \!+\! \varepsilon \delta \overline{X}_1
		\!+\! \dots \!+\! \varepsilon^{n-1} \delta \overline{X}_{n-1}
		\!+\! \varepsilon^{n} \delta \overline{X}_{n} ]
	\,\Big|_{\varepsilon=0} \biggr)
	= 0 \, .
\end{equation}
This essentially completes the proof by induction of the action~(\ref{nAction})
at arbitrary order.

%%%%%%%%%%%%%%%%%%%%%%%%%%%%%%%%%%%%%%%%%%%%%%%
%%%%%%%%%%%%%%%%%%%%%%%%%%%%%%%%%%%%%%%%%%%%%%%
%%%%%%%%%%%%%%%%%%%%%%%%%%%%%%%%%%%%%%%%%%%%%%%
%%%	REFERENCES
%%%%%%%%%%%%%%%%%%%%%%%%%%%%%%%%%%%%%%%%%%%%%%%
%%%%%%%%%%%%%%%%%%%%%%%%%%%%%%%%%%%%%%%%%%%%%%%
%%%%%%%%%%%%%%%%%%%%%%%%%%%%%%%%%%%%%%%%%%%%%%%

\end{document}